\documentclass[11pt]{article}
\usepackage{jheppub}
\usepackage[utf8]{inputenc}
\usepackage{amsmath}
\usepackage{tikz}
\usepackage{tkz-euclide}
\usepackage{dsfont}
\usepackage{amssymb}
\usepackage{graphicx}
\usepackage{mathrsfs}
\usepackage{appendix}
\usepackage{bbold}
\usepackage{color}

\def\ie{\begin{equation}\begin{aligned}}
\def\fe{\end{aligned}\end{equation}}

\begin{document}

\title{\textbf{Finite-cutoff holography and quasilocal thermodynamics of BTZ black holes in a cavity}}

\author[a]{Nazir A. Ganaie}

\affiliation[a]{Department of Physics, National Institute of Technology Srinagar, Srinagar, India-190006}

\emailAdd{nazirahmadgan.82225@jk.gov.in}

\abstract{
We show that the BTZ black hole in a finite radial cavity realizes a closed finite-cutoff thermodynamic system whose renormalized Brown-York stress tensor is simultaneously the quasilocal stress tensor of the cavity and the stress tensor of the cutoff dual theory. The central result is the exact finite-radius Hamilton-Jacobi equation for the wall stress tensor and its realization by the static and rotating BTZ families in local wall variables. In the homogeneous wall frame this equation becomes a nonlinear equation of state, with the corresponding gravitational deformation parameter fixed entirely by the bulk couplings. We use this structure to formulate the rotating cavity directly in terms of wall-measured grand-canonical data, derive the corresponding quasilocal first law and radial flow equations, and identify the finite-size Hawking-Page transition. The same wall dictionary gives an exact finite-cutoff spectral map and a deformed Cardy density of states in quasilocal energy variables. The cavity radius therefore plays a controlled double role: it is the physical size of the Brown-York thermodynamic system and the cutoff scale of the dual finite-radius theory.
}

\maketitle
\newpage

\section{Introduction}

Black-hole thermodynamics acquired a precise semiclassical footing when horizon area, temperature, and Euclidean regularity were understood as components of a single gravitational thermodynamic framework, so that the gravitational path integral with fixed boundary data defines genuine equilibrium ensembles rather than a formal analogy \cite{Bekenstein1973,BardeenCarterHawking1973,Hawking1975,GibbonsHawking1977,York1986,WhitingYork1988,BrownYork1993,HawkingPage1983}. Within this framework, finite-boundary ensembles are especially valuable because they render the thermodynamic control parameters local and operational: one fixes the intrinsic geometry of a timelike wall, interprets that wall as the thermal environment, and obtains quasilocal energy and mechanical response variables directly from the renormalized Dirichlet variational principle \cite{York1986,WhitingYork1988,BrownComerMartinezMelmedWhitingYork1990,BrownYork1993,BalasubramanianKraus1999,deHaroSkenderisSolodukhin2001}. In anti-de Sitter gravity this viewpoint acquires additional significance because radial position is simultaneously a geometric datum in the bulk and a scale variable in the dual description \cite{Maldacena1998,GubserKlebanovPolyakov1998,Witten1998AdSCFT,Skenderis2002}. 

In three dimensions these ideas admit an especially rigid realization. Einstein gravity with negative cosmological constant has no local propagating graviton, the BTZ family provides an exact black-hole laboratory with controlled Euclidean thermodynamics and conserved charges, and the asymptotic theory is governed by the Brown-Henneaux Virasoro symmetry with central charge fixed exactly \cite{BanadosTeitelboimZanelli1992,BanadosHenneauxTeitelboimZanelli1993,BrownHenneaux1986,Carlip1995,Carlip1998}. The same setting also supports a finite-cutoff interpretation through holographic renormalization and, in two-dimensional language, through solvable irrelevant deformations of \(T\bar T\) type, for which the stress tensor obeys an exact quadratic trace-flow equation and the finite-volume spectrum takes a universal square-root form \cite{BalasubramanianKraus1999,deHaroSkenderisSolodukhin2001,Zamolodchikov2004,CavagliaNegroSzecsenyiTateo2016,McGoughMezeiVerlinde2018,GiveonItzhakiKutasov2017,SmirnovZamolodchikov2017,Shyam2017}.

The central claim of this paper is that the BTZ cavity furnishes an exact finite-cutoff system whose quasilocal thermodynamics, holographic radial flow, and deformed two-dimensional interpretation are carried by one and the same renormalized wall stress tensor. More precisely, we treat a circular cavity at radius \(R\) as a genuine finite holographic screen and show that the renormalized Brown-York tensor on that screen satisfies the exact finite-radius Hamilton-Jacobi identity
\begin{equation}
T^{i}{}_{i}
=
\frac{\ell}{16\pi G}R[h]
+
4\pi G\ell\!\left(T_{ij}T^{ij}-\bigl(T^{i}{}_{i}\bigr)^{2}\right),
\label{eq:intro_trace_relation}
\end{equation}
which reduces for stationary homogeneous wall states on the flat BTZ wall cylinder to a nonlinear equation of state \cite{BalasubramanianKraus1999,deHaroSkenderisSolodukhin2001,Skenderis2002,McGoughMezeiVerlinde2018,Shyam2017}. This relation is the structural core of the analysis. It implies that the cavity wall supports a closed thermodynamic description at finite radius, with local observables determined intrinsically on the wall and with radial motion of the cutoff surface acting as an exact flow of those observables. The finite cavity therefore constitutes a quasilocal gravitational system in the Brown-York sense and, simultaneously, an explicit realization of finite-cutoff AdS$_3$/CFT$_2$ with a gravitationally fixed deformation parameter.

The BTZ family then provides the complete analytic realization of this structure. For static and rotating geometries we express the local wall energy, momentum density, pressure, temperature, and angular potential directly in terms of the bulk parameters $(r_{+},r_{-})$ and show that they satisfy the exact quasilocal first law with the wall circumference $L=2\pi R$ as a genuine thermodynamic variable. In the Euclidean theory the wall torus determines the allowed smooth bulk fillings, so the cavity ensemble admits a precise comparison between BTZ and thermal AdS\(_3\) saddles. One consequence is that the finite-size Hawking-Page transition occurs at
\begin{equation}
T_{c}(R)=\frac{1}{2\pi R},
\label{eq:intro_Tc}
\end{equation}
which depends only on the proper wall radius and therefore has a direct interpretation both as a quasilocal phase boundary and as the transition scale of the cutoff theory on a circle \cite{HawkingPage1983,Witten1998Thermal,HuangTao2022}.

\paragraph{Main results.}
The results needed for the rest of the paper can be summarized as follows. First, the renormalized wall stress tensor obeys the local finite-cutoff identity \eqref{eq:intro_trace_relation}, and hence the flat-wall BTZ equation of state
\begin{equation*}
    p-\epsilon=8\pi G\ell(\epsilon p-j^2).
\end{equation*}
Second, the rotating BTZ cavity admits a closed wall-variable description in terms of
\begin{equation*}
    (r_+,r_-)
\longmapsto
(\epsilon,j,p,E,J,T_R,\Omega_R),
\end{equation*}
with the exact first law
\[
dE=T_R dS+\Omega_R dJ-PdL,
\qquad L=2\pi R.
\]
Third, the static Euclidean saddle competition gives the finite-size Hawking-Page scale
\[
T_c(R)=\frac{1}{2\pi R},
\]
which is equivalently the point at which the proper thermal and spatial cycles of the wall torus have equal length. Fourth, radial displacement of the wall at fixed bulk state gives an exact flow,
\[
R\left(\frac{\partial \epsilon}{\partial R}\right)_{r_+,r_-}=-(\epsilon+p),
\qquad
R\left(\frac{\partial j}{\partial R}\right)_{r_+,r_-}=-2j.
\]
Finally, the wall energy is related algebraically to the ultraviolet BTZ charges, giving in the static sector
\[
M=\frac{R}{\ell}E-2GE^2,
\qquad
S(E,R)=2\pi\sqrt{\frac{c}{3}\left(RE-2G\ell E^2\right)}.
\]
These formulas are the finite-radius dictionary that distinguishes the present construction from a review of previously separate BTZ-cavity, Brown-York, and \(T\bar T\) ingredients \cite{BrownYork1993,HuangTao2022,McGoughMezeiVerlinde2018,CavagliaNegroSzecsenyiTateo2016}.

The later sections develop this claim in three layers. We first isolate the universal finite-cutoff stress-tensor relation and the corresponding radial flow law in a form independent of any explicit BTZ metric function. We then specialize to the BTZ cavity and derive the exact static and rotating wall thermodynamics, the off-shell free-energy landscape, and the finite-size Hawking-Page transition. We finally interpret the same wall system as a finite-cutoff holographic theory, derive the exact dictionary between wall variables and deformed two-dimensional thermodynamic data, and use that dictionary to formulate the finite-cutoff density of states together with the leading quantum corrections around smooth saddles. The resulting picture is that the cavity radius plays a double role with exact control: it is the physical size of the quasilocal thermodynamic system in the bulk and the renormalization scale of the dual cutoff description.

A recurring issue in finite-radius gravity is the normalization of the wall energy and stress tensor. At finite cutoff, the choice of local counterterm or subtraction prescription is part of the definition of the ensemble itself, because it fixes the vacuum energy of the wall theory. We work throughout with the standard renormalized Dirichlet action in AdS\(_3\), normalized so that massless BTZ carries vanishing quasilocal energy \cite{BalasubramanianKraus1999,deHaroSkenderisSolodukhin2001,Skenderis2002}. With this convention, the Brown-York tensor, the Euclidean saddle-point action, the radial Hamilton-Jacobi flow, and the finite-cutoff holographic interpretation fit into one coherent framework.

\section{Relation to previous BTZ cavity and finite-cutoff analyses}

A precise novelty statement for the present problem requires that three pre-existing lines of work be separated conceptually before they are recombined. The first line is the Brown-York quasilocal formalism and its AdS-renormalized extension, in which a finite timelike boundary carries a conserved stress tensor obtained from the renormalized Dirichlet action \cite{York1986,WhitingYork1988,BrownYork1993,BalasubramanianKraus1999,deHaroSkenderisSolodukhin2001,Skenderis2002}. The second line is the thermodynamics of BTZ black holes in a finite cavity, where one fixes local wall data and studies the resulting canonical or grand-canonical ensemble of smooth Euclidean fillings \cite{CarlipTeitelboim1995,BradenBrownWhitingYork1990,HuangTao2022}. The third line is the finite-cutoff AdS\(_3\)/CFT\(_2\) and \(T\bar T\) literature, in which a Dirichlet wall at finite radius is interpreted as the scale of an irrelevant deformation of a two-dimensional theory and the stress tensor obeys a quadratic trace-flow law \cite{Zamolodchikov2004,CavagliaNegroSzecsenyiTateo2016,McGoughMezeiVerlinde2018,GiveonItzhakiKutasov2017,SmirnovZamolodchikov2017,HartmanKruthoffShaghoulianTajdini2019,Taylor2018,Shyam2017}. The present paper belongs to the intersection of these three subjects, and its contribution must therefore be formulated at that intersection rather than by comparison with any one ingredient in isolation.

The Brown-York part of the story is classical and well established \cite{York1986,WhitingYork1988,BrownComerMartinezMelmedWhitingYork1990,BrownYork1993}. For a timelike wall \(\Sigma_R\) with induced metric \(h_{ij}\), outward unit normal \(n^\mu\), and renormalized Dirichlet action
\begin{equation}
S_{\mathrm{ren}}[g;R]
=
\frac{1}{16\pi G}\int_M d^3x\,\sqrt{-g}\left(\mathcal{R}[g]+\frac{2}{\ell^2}\right)
-\frac{1}{8\pi G}\int_{\Sigma_R} d^2x\,\sqrt{-h}\,K
-\frac{1}{8\pi G\ell}\int_{\Sigma_R} d^2x\,\sqrt{-h},
\label{eq:sec2_Sren}
\end{equation}
the renormalized wall stress tensor is
\begin{equation}
T^{\mathrm{ren}}_{ij}
=
\frac{1}{8\pi G}\left(Kh_{ij}-K_{ij}-\frac{1}{\ell}h_{ij}\right),
\label{eq:sec2_BYtensor}
\end{equation}
and obeys the conservation law
\begin{equation}
D_i T_{\mathrm{ren}}^{\,i}{}_{j}=0.
\label{eq:sec2_BYcons}
\end{equation}
For every Killing field \(\xi^i\) of the wall geometry one obtains a conserved Brown-York charge
\begin{equation}
Q_\xi
=
\int_{C_R} d\phi\,\sqrt{\sigma}\,u_i\,\xi_j\,T_{\mathrm{ren}}^{\,ij},
\label{eq:sec2_BYcharge}
\end{equation}
with \(u^i\) the wall observer velocity and \(C_R\) the spatial wall circle. These formulas provide the correct quasilocal language and, after AdS renormalization, the standard finite-radius holographic stress tensor technology \cite{BrownYork1993,BalasubramanianKraus1999,deHaroSkenderisSolodukhin2001,Skenderis2002}. By themselves, however, they do not determine the BTZ cavity ensemble, the Euclidean saddle competition at fixed wall torus, the finite-size Hawking-Page scale, or the exact finite-cutoff dictionary in wall thermodynamic variables.

The BTZ cavity literature contributes the second ingredient. For a wall at radius \(R\), the natural intensive variables are the local inverse temperature and wall angular potential,
\begin{equation}
\beta_R=\beta_H N(R), \qquad
\Omega_R=\frac{\Omega_H-\Omega(R)}{N(R)},
\label{eq:sec2_localpotentials}
\end{equation}
so the ensemble is defined by finite wall data rather than by asymptotic quantities. In particular, the static branch exhibits a Hawking-Page type transition controlled by the cavity size, while the rotating branch admits a grand-canonical local stability analysis in wall variables \cite{BradenBrownWhitingYork1990,HuangTao2022}. Thus the existence of a cavity-controlled transition and the existence of rotating BTZ cavity thermodynamics are not by themselves new features of the present work. The genuinely nontrivial question is whether the exact wall thermodynamic variables of the BTZ cavity can be organized as the stress-tensor data of a finite-cutoff theory with no extra phenomenological input. The present paper answers that question affirmatively.

The finite-cutoff AdS\(_3\)/CFT\(_2\) and \(T\bar T\) literature contributes the third ingredient. In that framework, a Dirichlet wall at finite radius is interpreted as the locus on which the stress tensor of the deformed theory is evaluated, and the Hamilton-Jacobi constraint implies a quadratic trace relation of the form
\begin{equation}
T^{i}{}_{i}
=
\frac{\ell}{16\pi G}R[h]
+
4\pi G\ell\left(T_{ij}T^{ij}-\bigl(T^{i}{}_{i}\bigr)^2\right),
\qquad
\mu_{\mathrm{grav}}=8\pi G\ell.
\label{eq:sec2_traceflow}
\end{equation}
On a flat wall cylinder this reduces to
\begin{equation}
p-\epsilon
=
\mu_{\mathrm{grav}}\bigl(\epsilon p-j^2\bigr),
\label{eq:sec2_flattraceflow}
\end{equation}
with \(\epsilon\), \(j\), and \(p\) the local energy density, momentum density, and pressure. The appearance of such a quadratic stress-tensor relation in cutoff AdS\(_3\) holography and in \(T\bar T\)-deformed field theory is already part of the established theoretical framework \cite{Zamolodchikov2004,CavagliaNegroSzecsenyiTateo2016,McGoughMezeiVerlinde2018,GiveonItzhakiKutasov2017,SmirnovZamolodchikov2017,HartmanKruthoffShaghoulianTajdini2019,Taylor2018,Shyam2017}. What is absent from the earlier BTZ cavity literature is the exact realization of this flow law directly in terms of the renormalized BTZ wall observables and their Euclidean thermodynamic potentials.

The present paper fills that gap by constructing the finite-radius map
\begin{equation}
(r_+,r_-)
\longmapsto
\bigl(
\epsilon(R),\,j(R),\,p(R),\,E(R),\,J,\,T_R,\,\Omega_R
\bigr),
\label{eq:sec2_map}
\end{equation}
with
\begin{equation}
\epsilon(R)
=
\frac{1}{8\pi G\ell}
\left(
1-\frac{\sqrt{(R^2-r_+^2)(R^2-r_-^2)}}{R^2}
\right),
\qquad
j(R)
=
\frac{r_+r_-}{8\pi G\ell R^2},
\label{eq:sec2_epsj}
\end{equation}
\begin{equation}
p(R)
=
\frac{1}{8\pi G\ell}
\left(
\frac{R^4-r_+^2r_-^2}{R^2\sqrt{(R^2-r_+^2)(R^2-r_-^2)}}-1
\right),
\label{eq:sec2_p}
\end{equation}
\begin{equation}
E(R)=2\pi R\,\epsilon(R),
\qquad
J=2\pi R^2 j(R)=\frac{r_+r_-}{4G\ell},
\label{eq:sec2_EJ}
\end{equation}
together with the exact wall potentials
\begin{equation}
T_R
=
\frac{R(r_+^2-r_-^2)}{2\pi \ell r_+\sqrt{(R^2-r_+^2)(R^2-r_-^2)}},
\qquad
\Omega_R
=
\frac{r_-\sqrt{R^2-r_+^2}}{r_+R\sqrt{R^2-r_-^2}}.
\label{eq:sec2_TROR}
\end{equation}
These quantities satisfy simultaneously the nonlinear local equation of state \eqref{eq:sec2_flattraceflow} and the quasilocal first law
\begin{equation}
dE=T_R\,dS+\Omega_R\,dJ-P\,dL,
\qquad
L=2\pi R,
\label{eq:sec2_firstlaw}
\end{equation}
with radial evolution governed by
\begin{equation}
R\left(\frac{\partial \epsilon}{\partial R}\right)_{r_+,r_-}
=
-(\epsilon+p).
\label{eq:sec2_rflow}
\end{equation}
The significance of these formulas is that the cavity variables close on an exact finite-dimensional thermodynamic system whose stress tensor is at once Brown-York, quasilocal, and finite-cutoff holographic. This goes beyond the mere existence of quasilocal charges, beyond the mere existence of BTZ cavity thermodynamics, and beyond the abstract identification of cutoff AdS\(_3\) with \(T\bar T\)-type flow.

The same statement extends to the microscopic sector. The wall energy is related exactly to the asymptotic BTZ mass parameter by
\begin{equation}
M=\frac{R}{\ell}E-2GE^2,
\label{eq:sec2_MErelation}
\end{equation}
and the ultraviolet Cardy degeneracy then induces the finite-cutoff entropy formula
\begin{equation}
S(E,R)
=
2\pi
\sqrt{
\frac{c}{3}\left(RE-2G\ell E^2\right)
},
\qquad
c=\frac{3\ell}{2G}.
\label{eq:sec2_deformedCardy}
\end{equation}
Equations \eqref{eq:sec2_MErelation} and \eqref{eq:sec2_deformedCardy} are not consequences of the Brown-York tensor alone, nor of the cavity first law alone, nor of the abstract cutoff correspondence alone. They arise only after the exact wall thermodynamics has been identified with the finite-cutoff stress-tensor data and then carried into the state-counting problem \cite{Cardy1986,Strominger1998,McGoughMezeiVerlinde2018}.

The closest overlap in the BTZ cavity literature is the 2022 analysis of Huang and Tao, which already studied static, rotating, and charged BTZ black holes in a cavity, derived thermodynamic quantities from the Euclidean action, and analyzed local stability and phase structure \cite{HuangTao2022}. The present work should therefore not be read as claiming novelty for the existence of a rotating cavity ensemble, for the positivity of static or rotating heat capacities in isolation, or for the occurrence of a cavity-controlled Hawking-Page transition as such. Its distinct contribution is the exact renormalized wall formulation in which the BTZ cavity is shown to realize one closed finite-cutoff system with a universal local trace-flow law, an exact wall-based radial RG flow, an explicit gravitational value of the deformation parameter, and a microscopic density-of-states map written directly in the same wall variables.

The correct positioning of the paper is therefore narrow and structural. It does not introduce Brown-York quasilocal thermodynamics, it does not introduce BTZ cavity thermodynamics, and it does not introduce finite-cutoff AdS\(_3\)/CFT\(_2\) or \(T\bar T\) flow. It shows that, for the BTZ cavity, these three frameworks are realized by the same renormalized wall system with exact local, thermodynamic, and microscopic formulas. The remainder of the paper develops this claim by deriving the universal stress-tensor flow law, evaluating it on the BTZ family, analyzing the resulting phase structure, and identifying the same wall variables with the observables of the cutoff dual theory.

\begin{figure}[t]
  \centering
  \includegraphics[width=1.0\linewidth]{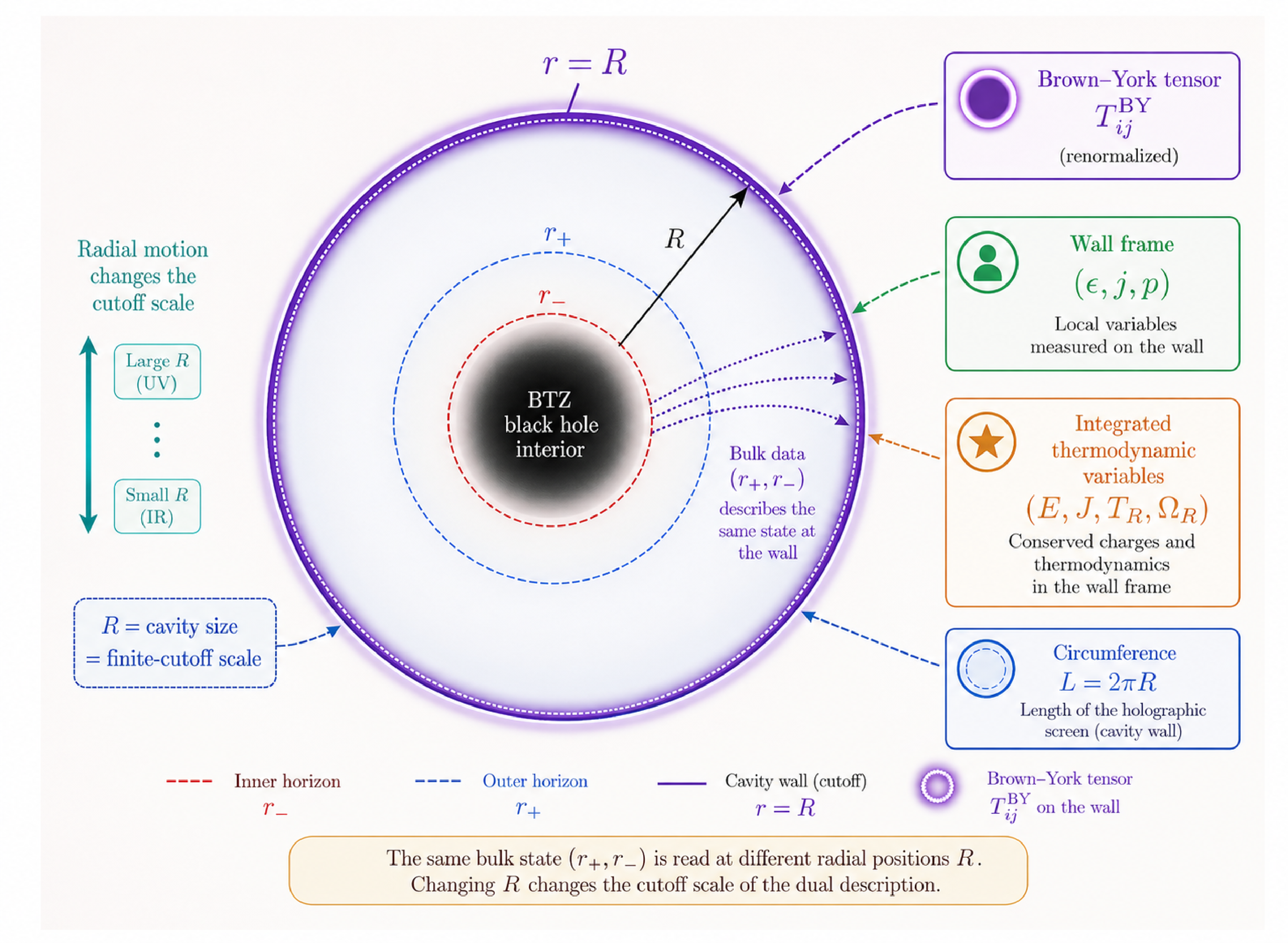}
  \caption{
  Finite-radius BTZ cavity as a holographic screen. The bulk state is labelled
  by the horizon data \((r_{+},r_{-})\), while the wall at \(r=R\) carries the
  renormalized Brown-York tensor \(T^{\rm BY}_{ij}\). In the wall frame this
  tensor defines the local data \((\epsilon,j,p)\), the integrated thermodynamic
  variables \((E,J,T_R,\Omega_R)\), and the circumference \(L=2\pi R\). Radial
  motion of the wall changes the scale at which the same bulk state is described,
  so \(R\) is both the cavity size and the finite-cutoff scale.
  }
  \label{fig:btz-cavity-screen}
\end{figure}

Figure~\ref{fig:btz-cavity-screen} summarizes the finite-radius dictionary used throughout the paper: the same wall stress tensor supplies the Brown-York quasilocal variables and the finite-cutoff holographic stress tensor.

\section{Universal finite-cutoff equation of state and radial flow}
\label{sec:section_3}

The finite-cutoff formulation admits a structural statement that precedes the explicit evaluation on any particular BTZ saddle. In three-dimensional Einstein gravity with negative cosmological constant, the absence of local propagating bulk gravitons implies that the radial dependence of the on-shell action is governed entirely by the Hamiltonian and momentum constraints. Once the theory is placed on a timelike Dirichlet wall and renormalized by the standard local AdS\(_3\) counterterm, these constraints descend to an exact local relation obeyed by the wall stress tensor. This relation is the precise point at which quasilocal thermodynamics, radial Hamilton-Jacobi flow, and the finite-cutoff \(T\bar T\)-type structure meet in a single equation \cite{BrownYork1993,BalasubramanianKraus1999,deHaroSkenderisSolodukhin2001,Skenderis2002,Zamolodchikov2004,CavagliaNegroSzecsenyiTateo2016,McGoughMezeiVerlinde2018,GiveonItzhakiKutasov2017,Taylor2018,Shyam2017}. The purpose of the present section is to isolate that universal content in a form that depends only on the bulk field equations, the finite-radius variational principle, and the intrinsic geometry of the wall. All BTZ-specific formulas derived later are then exact realizations of the identities established here.

Consider three-dimensional Einstein gravity with cosmological constant \(\Lambda=-\ell^{-2}\) on a manifold \(M\) with timelike boundary \(\Sigma\), outward-pointing unit normal \(n^\mu\), induced metric \(h_{ij}\), and extrinsic curvature
\begin{equation}
K_{ij}=\frac{1}{2}\,\mathcal{L}_{n}h_{ij}.
\label{eq:sec3_Kij}
\end{equation}
The renormalized Dirichlet action is
\begin{equation}
S_{\mathrm{ren}}[g;\Sigma]
=
\frac{1}{16\pi G}\int_{M} d^{3}x\,\sqrt{-g}\left(\mathcal{R}[g]+\frac{2}{\ell^{2}}\right)
-\frac{1}{8\pi G}\int_{\Sigma} d^{2}x\,\sqrt{-h}\,K
-\frac{1}{8\pi G\ell}\int_{\Sigma} d^{2}x\,\sqrt{-h}.
\label{eq:sec3_Sren}
\end{equation}
Its variation takes the standard form
\begin{equation}
\delta S_{\mathrm{ren}}
=
\frac{1}{16\pi G}\int_{M} d^{3}x\,\sqrt{-g}
\left(
R_{\mu\nu}-\frac{1}{2}\mathcal{R}\,g_{\mu\nu}-\frac{1}{\ell^{2}}g_{\mu\nu}
\right)\delta g^{\mu\nu}
+
\frac{1}{2}\int_{\Sigma} d^{2}x\,\sqrt{-h}\,T_{ij}\,\delta h^{ij},
\label{eq:sec3_varSren}
\end{equation}
where the renormalized Brown--York tensor is
\begin{equation}
T_{ij}
=
\frac{1}{8\pi G}
\left(
Kh_{ij}-K_{ij}-\frac{1}{\ell}h_{ij}
\right).
\label{eq:sec3_BY}
\end{equation}
On shell the bulk equations are
\begin{equation}
R_{\mu\nu}=-\frac{2}{\ell^{2}}g_{\mu\nu},
\qquad
\mathcal R[g]=-\frac{6}{\ell^{2}}.
\label{eq:sec3_Einstein}
\end{equation}
The momentum constraint is the contracted Codazzi equation,
\begin{equation}
D_{i}\!\left(K^{i}{}_{j}-K\delta^{i}{}_{j}\right)=0,
\label{eq:sec3_Codazzi}
\end{equation}
and after substituting \eqref{eq:sec3_BY} it becomes the wall conservation law
\begin{equation}
D_{i}T^{i}{}_{j}=0.
\label{eq:sec3_conservation}
\end{equation}
The Hamiltonian constraint follows from the Gauss relation for the timelike hypersurface \(\Sigma\),
\begin{equation}
R[h]
=
\mathcal R[g]-2R_{\mu\nu}n^{\mu}n^{\nu}
+
K^{2}-K_{ij}K^{ij}.
\label{eq:sec3_Gauss}
\end{equation}
Using \eqref{eq:sec3_Einstein} and \(g_{\mu\nu}n^{\mu}n^{\nu}=1\), one finds
\begin{equation}
R[h]
=
-\frac{2}{\ell^{2}}
+
K^{2}-K_{ij}K^{ij},
\label{eq:sec3_Rh}
\end{equation}
or equivalently
\begin{equation}
K_{ij}K^{ij}-K^{2}
=
-R[h]-\frac{2}{\ell^{2}}.
\label{eq:sec3_Hconstraint}
\end{equation}
To rewrite this identity entirely in wall variables, define
\begin{equation}
\tau_{ij}:=8\pi G\,T_{ij}
=
Kh_{ij}-K_{ij}-\frac{1}{\ell}h_{ij}.
\label{eq:sec3_tauij}
\end{equation}
Since the wall is two-dimensional, its trace is
\begin{equation}
\tau:=h^{ij}\tau_{ij}=K-\frac{2}{\ell},
\label{eq:sec3_tautrace}
\end{equation}
and therefore
\begin{equation}
K_{ij}
=
\left(\tau+\frac{1}{\ell}\right)h_{ij}-\tau_{ij}.
\label{eq:sec3_Kfromtau}
\end{equation}
The quadratic invariant appearing in \eqref{eq:sec3_Hconstraint} is then
\begin{align}
K_{ij}K^{ij}
&=
\left[\left(\tau+\frac{1}{\ell}\right)h_{ij}-\tau_{ij}\right]
\left[\left(\tau+\frac{1}{\ell}\right)h^{ij}-\tau^{ij}\right]
\nonumber\\
&=
2\left(\tau+\frac{1}{\ell}\right)^{2}
-2\left(\tau+\frac{1}{\ell}\right)\tau
+\tau_{ij}\tau^{ij}
\nonumber\\
&=
\tau_{ij}\tau^{ij}
+\frac{2\tau}{\ell}
+\frac{2}{\ell^{2}},
\label{eq:sec3_KijKij}
\end{align}
while
\begin{equation}
K^{2}
=
\left(\tau+\frac{2}{\ell}\right)^{2}
=
\tau^{2}+\frac{4\tau}{\ell}+\frac{4}{\ell^{2}}.
\label{eq:sec3_Ksq}
\end{equation}
Substituting \eqref{eq:sec3_KijKij} and \eqref{eq:sec3_Ksq} into \eqref{eq:sec3_Hconstraint} gives
\begin{equation}
\tau_{ij}\tau^{ij}-\tau^{2}-\frac{2\tau}{\ell}-\frac{2}{\ell^{2}}
=
-R[h]-\frac{2}{\ell^{2}},
\label{eq:sec3_tauconstraint1}
\end{equation}
hence
\begin{equation}
\tau
=
\frac{\ell}{2}\,R[h]
+
\frac{\ell}{2}\left(\tau_{ij}\tau^{ij}-\tau^{2}\right).
\label{eq:sec3_tauconstraint2}
\end{equation}
Returning to the Brown-York tensor by \(\tau_{ij}=8\pi G\,T_{ij}\), one obtains the exact finite-radius Hamilton-Jacobi identity
\begin{equation}
T^{i}{}_{i}
=
\frac{\ell}{16\pi G}\,R[h]
+
4\pi G\ell
\left(
T_{ij}T^{ij}
-
\bigl(T^{i}{}_{i}\bigr)^{2}
\right).
\label{eq:sec3_traceflow}
\end{equation}
Equation \eqref{eq:sec3_traceflow} is the central structural result of the finite-cutoff construction. It follows directly from the bulk equations of motion and the renormalized Dirichlet variational principle, and it is therefore independent of any particular choice of stationary solution. In the holographic interpretation it is the exact local stress-tensor flow law of the cutoff theory. In the quasilocal interpretation it is the intrinsic equation that controls the local thermodynamic response of the wall \cite{BalasubramanianKraus1999,deHaroSkenderisSolodukhin2001,McGoughMezeiVerlinde2018,Shyam2017}.

For a stationary homogeneous state on a wall cylinder, choose an orthonormal frame \((e_{\hat t},e_{\hat\phi})\) adapted to the wall observer and write the stress tensor as
\begin{equation}
T_{\hat a\hat b}
=
\begin{pmatrix}
\epsilon & -j\\
-j & p
\end{pmatrix},
\label{eq:sec3_Torth}
\end{equation}
where \(\epsilon\) is the local energy density, \(j\) is the momentum density along the circle, and \(p\) is the wall pressure. The scalar invariants are
\begin{equation}
T^{\hat a}{}_{\hat a}=p-\epsilon,
\qquad
T_{\hat a\hat b}T^{\hat a\hat b}
=
\epsilon^{2}+p^{2}-2j^{2}.
\label{eq:sec3_invariants}
\end{equation}
Substitution into \eqref{eq:sec3_traceflow} yields
\begin{equation}
p-\epsilon
=
\frac{\ell}{16\pi G}\,R[h]
+
8\pi G\ell\left(\epsilon p-j^{2}\right).
\label{eq:sec3_eos_curved}
\end{equation}
For the flat wall cylinder relevant to the BTZ cavity one has \(R[h]=0\), and the equation of state reduces to
\begin{equation}
p-\epsilon
=
8\pi G\ell\left(\epsilon p-j^{2}\right).
\label{eq:sec3_eos_flat}
\end{equation}
This is the exact nonlinear equation of state of the homogeneous finite-cutoff wall fluid. It shows that at fixed cutoff the pressure is determined algebraically once the local energy density and momentum density are known. In the static sector \(j=0\), equation \eqref{eq:sec3_eos_flat} becomes
\begin{equation}
p
=
\frac{\epsilon}{1-8\pi G\ell\,\epsilon},
\label{eq:sec3_staticEOS}
\end{equation}
which exhibits the finite-radius departure from the ultraviolet conformal relation \(p=\epsilon\). The same quadratic structure is the one that appears in the local \(T\bar T\) flow law, with the gravitational normalization
\begin{equation}
\mu_{\mathrm{grav}}=8\pi G\ell
\label{eq:sec3_mugrav}
\end{equation}
fixed uniquely by the bulk action and counterterm prescription \cite{Zamolodchikov2004,CavagliaNegroSzecsenyiTateo2016,McGoughMezeiVerlinde2018,Taylor2018,Shyam2017}.

The same universal framework yields the radial flow law. Let the proper circumference of the wall be
\begin{equation}
L=2\pi R,
\label{eq:sec3_L}
\end{equation}
and let \(E(L)\) denote the integrated quasilocal energy of a homogeneous stationary state. Homogeneity gives
\begin{equation}
E(L)=L\,\epsilon(L).
\label{eq:sec3_ELeps}
\end{equation}
The finite-radius first law takes the form
\begin{equation}
dE=T\,dS+\Omega\,dJ-p\,dL.
\label{eq:sec3_firstlaw}
\end{equation}
If the cutoff surface is moved while the underlying bulk state is kept fixed, then the state labels \(S\) and \(J\) remain fixed and \eqref{eq:sec3_firstlaw} implies
\begin{equation}
\left(\frac{\partial E}{\partial L}\right)_{S,J}
=
-p.
\label{eq:sec3_dEdL}
\end{equation}
Using \eqref{eq:sec3_ELeps}, one finds
\begin{equation}
\epsilon
+
L\left(\frac{\partial \epsilon}{\partial L}\right)_{S,J}
=
-p,
\label{eq:sec3_epsflowL1}
\end{equation}
hence
\begin{equation}
L\left(\frac{\partial \epsilon}{\partial L}\right)_{S,J}
=
-(\epsilon+p).
\label{eq:sec3_epsflowL2}
\end{equation}
Since \(L=2\pi R\), this becomes
\begin{equation}
R\left(\frac{\partial \epsilon}{\partial R}\right)_{S,J}
=
-(\epsilon+p).
\label{eq:sec3_epsflowR}
\end{equation}
For a homogeneous rotating state the integrated angular momentum is
\begin{equation}
J=L\,R\,j=2\pi R^{2}j,
\label{eq:sec3_Jhom}
\end{equation}
so conservation of \(J\) along radial motion gives
\begin{equation}
R\left(\frac{\partial j}{\partial R}\right)_{J}
=
-2j.
\label{eq:sec3_jflow}
\end{equation}
Equations \eqref{eq:sec3_eos_flat}, \eqref{eq:sec3_epsflowR}, and \eqref{eq:sec3_jflow} form a closed finite-cutoff flow system for the homogeneous wall data. The first relation is algebraic, and the remaining two are first-order radial evolution equations. Once the state is specified at one radius, the local wall thermodynamics is determined at every other radius by exact flow.

The content of the section may therefore be summarized in a compact form. The renormalized Brown-York tensor obeys the conservation law \eqref{eq:sec3_conservation} and the exact trace relation \eqref{eq:sec3_traceflow}; for homogeneous wall states these reduce to the nonlinear equation of state \eqref{eq:sec3_eos_flat} and the radial flow equations \eqref{eq:sec3_epsflowR} and \eqref{eq:sec3_jflow}. The finite-cutoff wall is thus an intrinsically defined thermodynamic system whose local observables are constrained entirely by the bulk Einstein equations and the finite-radius variational principle. The subsequent sections specialize this universal structure to the BTZ family, where the explicit metric data provide the exact map from \((r_{+},r_{-})\) to the wall observables, the Euclidean saddle analysis provides the corresponding free energies and phase structure, and the cutoff holographic interpretation identifies the same wall variables with those of the deformed dual theory.

\section{Conceptual framework: cavity ensembles as finite radial cutoffs}

We now specialize the universal wall-stress-tensor structure to the BTZ family \cite{BanadosTeitelboimZanelli1992,BanadosHenneauxTeitelboimZanelli1993}. The bulk state is labelled by \((r_+,r_-)\), while the cavity radius \(R\) selects the local thermodynamic frame in which that state is measured. The same geometry therefore carries a one-parameter family of Brown-York descriptions, and radial motion of the wall is the finite-cutoff flow of the local observables.

The rotating BTZ metric is
\begin{equation}
ds^{2}
=
-N^{2}(r)dt^{2}+\frac{dr^{2}}{N^{2}(r)}
+r^{2}\bigl(d\phi+N^{\phi}(r)dt\bigr)^{2},
\qquad
\phi\sim\phi+2\pi,
\label{eq:sec4_BTZmetric}
\end{equation}
with
\begin{equation}
N^{2}(r)=\frac{(r^{2}-r_{+}^{2})(r^{2}-r_{-}^{2})}{\ell^{2}r^{2}},
\qquad
N^{\phi}(r)=-\frac{r_{+}r_{-}}{\ell r^{2}},
\qquad
0\le r_-\le r_+<R.
\label{eq:sec4_lapse}
\end{equation}
The ADM charges and horizon data are
\begin{equation}
M=\frac{r_{+}^{2}+r_{-}^{2}}{8G\ell^{2}},
\qquad
J=\frac{r_{+}r_{-}}{4G\ell},
\qquad
\Omega_H=\frac{r_-}{\ell r_+},
\qquad
T_H=\frac{r_+^2-r_-^2}{2\pi\ell^2 r_+}.
\label{eq:sec4_ADM_horizon}
\end{equation}
Introduce
\begin{equation}
\Delta(r)=\sqrt{(r^{2}-r_{+}^{2})(r^{2}-r_{-}^{2})},
\qquad
\Delta_R=\Delta(R),
\qquad
\Omega(R)=-N^\phi(R)=\frac{r_+r_-}{\ell R^2}.
\label{eq:sec4_Delta}
\end{equation}
In the corotating coordinate \(\varphi=\phi-\Omega(R)t\), the induced wall metric is
\begin{equation}
h_{ij}dx^idx^j\big|_{\Sigma_R}
=
-N^2(R)dt^2+R^2d\varphi^2,
\qquad
N(R)=\frac{\Delta_R}{\ell R},
\qquad
R[h]=0.
\label{eq:sec4_wallmetric}
\end{equation}
The corresponding orthonormal frame is
\begin{equation}
e_{\hat t}=\frac{1}{N(R)}\bigl(\partial_t+\Omega(R)\partial_\phi\bigr),
\qquad
e_{\hat\varphi}=\frac{1}{R}\partial_\phi,
\label{eq:sec4_frame}
\end{equation}
and the wall stress tensor is homogeneous,
\begin{equation}
T_{\hat a\hat b}(R)=
\begin{pmatrix}
\epsilon(R)&-j(R)\\
-j(R)&p(R)
\end{pmatrix}.
\label{eq:sec4_Torth}
\end{equation}

Euclidean regularity fixes the local intensive variables. The smooth identification
\begin{equation}
(\tau,\phi)\sim(\tau+\beta_H,\phi-i\beta_H\Omega_H)
\label{eq:sec4_EuclidID}
\end{equation}
gives, in the wall frame,
\begin{equation}
\beta_R=\beta_H N(R)=\frac{2\pi\ell r_+\Delta_R}{R(r_+^2-r_-^2)},
\qquad
T_R=\frac{R(r_+^2-r_-^2)}{2\pi\ell r_+\Delta_R},
\label{eq:sec4_betaTR}
\end{equation}
\begin{equation}
\Omega_R=\frac{\Omega_H-\Omega(R)}{N(R)}
=
\frac{r_-\sqrt{R^2-r_+^2}}{r_+R\sqrt{R^2-r_-^2}}.
\label{eq:sec4_OR}
\end{equation}
The Brown-York tensor gives
\begin{equation}
\epsilon(R)=\frac{1}{8\pi G\ell}\left(1-\frac{\Delta_R}{R^2}\right),
\qquad
j(R)=\frac{r_+r_-}{8\pi G\ell R^2},
\label{eq:sec4_epsj}
\end{equation}
\begin{equation}
p(R)=\frac{1}{8\pi G\ell}\left(\frac{R^4-r_+^2r_-^2}{R^2\Delta_R}-1\right).
\label{eq:sec4_p}
\end{equation}
The integrated wall variables are
\begin{equation}
E(R)=2\pi R\epsilon(R)=\frac{R}{4G\ell}\left(1-\frac{\Delta_R}{R^2}\right),
\qquad
J(R)=2\pi R^2j(R)=\frac{r_+r_-}{4G\ell},
\qquad
L=2\pi R,
\label{eq:sec4_EJL}
\end{equation}
while
\begin{equation}
S=\frac{\pi r_+}{2G}.
\label{eq:sec4_S}
\end{equation}
These quantities satisfy the finite-radius first law
\begin{equation}
dE=T_RdS+\Omega_RdJ-PdL,
\qquad
P=p(R).
\label{eq:sec4_firstlaw}
\end{equation}
At fixed bulk state, radial motion of the wall gives
\begin{equation}
\left(\frac{\partial E}{\partial R}\right)_{r_+,r_-}=-2\pi P,
\qquad
R\left(\frac{\partial \epsilon}{\partial R}\right)_{r_+,r_-}=-(\epsilon+p),
\label{eq:sec4_radialflow}
\end{equation}
\begin{equation}
R\left(\frac{\partial T_R}{\partial R}\right)_{r_+,r_-}
=-T_R\frac{R^4-r_+^2r_-^2}{(R^2-r_+^2)(R^2-r_-^2)},
\label{eq:sec4_TRflow}
\end{equation}
\begin{equation}
R\left(\frac{\partial\Omega_R}{\partial R}\right)_{r_+,r_-}
=
\Omega_R\left[
\frac{R^2(r_+^2-r_-^2)}{(R^2-r_+^2)(R^2-r_-^2)}-1
\right].
\label{eq:sec4_ORflow}
\end{equation}

Two limiting regimes fix the physical interpretation. For \(R\gg \ell,r_+,r_-\),
\begin{equation}
\epsilon=p=\frac{r_+^2+r_-^2}{16\pi G\ell R^2}+O(R^{-4}),
\qquad
j=\frac{r_+r_-}{8\pi G\ell R^2},
\label{eq:sec4_UV}
\end{equation}
so the wall stress tensor becomes asymptotically traceless; in the static sector \(\epsilon=p=(\pi c/6)T_R^2+O(R^{-4})\), with \(c=3\ell/(2G)\). Near the wall, \(R=r_++\delta\),
\begin{equation}
T_R\sim\frac{\sqrt{r_+^2-r_-^2}}{2\pi\ell\sqrt{2r_+\delta}},
\qquad
P\sim\frac{\sqrt{r_+^2-r_-^2}}{8\pi G\ell\sqrt{2r_+\delta}},
\qquad
E\to\frac{r_+}{4G\ell}.
\label{eq:sec4_nearwall}
\end{equation}
The local intensive variables diverge by redshift, while the integrated quasilocal energy remains finite. The BTZ cavity is therefore a finite thermodynamic screen: \((r_+,r_-)\) specify the state, and \(R\) specifies the scale at which the wall theory describes it.

\section{Euclidean formulation of static BTZ at finite cutoff}

The static finite-cutoff ensemble is defined by fixing the wall torus \((\beta_R,L)\), with \(L=2\pi R\), and summing over smooth Euclidean fillings \cite{WhitingYork1988,HawkingPage1983,HuangTao2022}. In the circularly symmetric sector the only relevant smooth fillings are Euclidean BTZ, where the thermal cycle contracts, and thermal AdS\(_3\), where the angular cycle contracts. The standard action evaluation is placed in Appendix~\ref{app:static_action}; this section records the resulting wall thermodynamics and saddle competition.

The Euclidean Dirichlet action is
\begin{equation}
I_E[g;R]
=
-\frac{1}{16\pi G}\int_M d^3x\sqrt g\left(\mathcal R[g]+\frac{2}{\ell^2}\right)
-\frac{1}{8\pi G}\int_{\partial M}d^2x\sqrt h\,K
+\frac{1}{8\pi G\ell}\int_{\partial M}d^2x\sqrt h.
\label{eq:sec5_IE}
\end{equation}
The static ansatz
\begin{equation}
ds^2=N^2(r)d\tau^2+\frac{dr^2}{F(r)}+r^2d\phi^2,
\qquad
\phi\sim\phi+2\pi,
\qquad
\tau\sim\tau+\beta,
\label{eq:sec5_ansatz}
\end{equation}
obeys the Einstein equations through
\begin{equation}
N(r)=\sqrt{F(r)},
\qquad
F(r)=\frac{r^2}{\ell^2}+c_0.
\label{eq:sec5_static_solution}
\end{equation}
Thus the two static fillings are
\begin{equation}
ds^2_{\rm BTZ}=f(r)d\tau^2+\frac{dr^2}{f(r)}+r^2d\phi^2,
\qquad
f(r)=\frac{r^2-r_+^2}{\ell^2},
\qquad
r\in[r_+,R],
\label{eq:sec5_BTZ}
\end{equation}
and
\begin{equation}
ds^2_{\rm AdS}=f_0(r)d\tau^2+\frac{dr^2}{f_0(r)}+r^2d\phi^2,
\qquad
f_0(r)=1+\frac{r^2}{\ell^2},
\qquad
r\in[0,R].
\label{eq:sec5_tAdS}
\end{equation}
Regularity of the BTZ bolt fixes
\begin{equation}
\beta_H=\frac{2\pi\ell^2}{r_+},
\qquad
T_H=\frac{r_+}{2\pi\ell^2},
\label{eq:sec5_betaH}
\end{equation}
and the proper wall temperature is
\begin{equation}
\beta_R=\beta_H\sqrt{f(R)}=\frac{2\pi\ell\sqrt{R^2-r_+^2}}{r_+},
\qquad
T_R=\frac{r_+}{2\pi\ell\sqrt{R^2-r_+^2}}.
\label{eq:sec5_TR}
\end{equation}
The map \(r_+\mapsto T_R\) is monotone, and its inverse is
\begin{equation}
r_+(R,T_R)=\frac{2\pi\ell RT_R}{\sqrt{1+4\pi^2\ell^2T_R^2}},
\qquad
\sqrt{R^2-r_+^2}=\frac{R}{\sqrt{1+4\pi^2\ell^2T_R^2}}.
\label{eq:sec5_rplusinvert}
\end{equation}

The on-shell actions are
\begin{equation}
I_{\rm BH}(R,r_+)=\frac{\pi R}{2Gr_+}\left(\sqrt{R^2-r_+^2}-R\right),
\label{eq:sec5_IBH_rplus}
\end{equation}
\begin{equation}
I_{\rm AdS}(R,\beta_R)=\frac{\beta_R}{4G\ell}\left(R-\sqrt{R^2+\ell^2}\right).
\label{eq:sec5_IAdS}
\end{equation}
Equivalently,
\begin{equation}
F_{\rm BH}(R,T_R)=\frac{R}{4G\ell}\left(1-\sqrt{1+4\pi^2\ell^2T_R^2}\right),
\label{eq:sec5_FBH}
\end{equation}
\begin{equation}
F_{\rm AdS}(R)=\frac{1}{4G\ell}\left(R-\sqrt{R^2+\ell^2}\right),
\label{eq:sec5_FAdS}
\end{equation}
and therefore
\begin{equation}
F_{\rm BH}-F_{\rm AdS}
=\frac{1}{4G\ell}\left(\sqrt{R^2+\ell^2}-R\sqrt{1+4\pi^2\ell^2T_R^2}\right).
\label{eq:sec5_DeltaF}
\end{equation}
The Brown-York energy follows from the canonical derivative,
\begin{equation}
E_{\rm BH}=\left(\frac{\partial I_{\rm BH}}{\partial\beta_R}\right)_R
=
\frac{R}{4G\ell}\left(1-\frac{1}{\sqrt{1+4\pi^2\ell^2T_R^2}}\right)
=
\frac{R-\sqrt{R^2-r_+^2}}{4G\ell}.
\label{eq:sec5_Echeck}
\end{equation}
The Euclidean and Brown-York constructions therefore give the same finite-radius internal energy \cite{BrownYork1993,BalasubramanianKraus1999}.

The useful limits are compact. For \(2\pi\ell T_R\ll1\),
\begin{equation}
F_{\rm BH}=-\frac{\pi^2\ell R}{2G}T_R^2+\frac{\pi^4\ell^3R}{2G}T_R^4+O(T_R^6).
\label{eq:sec5_lowT}
\end{equation}
For \(R\gg\ell\) at fixed \(r_+\),
\begin{equation}
T_R=\frac{r_+}{2\pi\ell R}+O(R^{-3}),
\qquad
F_{\rm BH}=-\frac{r_+^2}{8G\ell R}+O(R^{-3}),
\qquad
F_{\rm AdS}=-\frac{\ell}{8GR}+O(R^{-3}).
\label{eq:sec5_largeR}
\end{equation}
Near the wall, \(r_+\to R^-\), the local temperature and \(|F_{\rm BH}|\) diverge by blueshift. The phase transition obtained later is the exact crossing of \eqref{eq:sec5_FBH} and \eqref{eq:sec5_FAdS}.

\section{Quasilocal thermodynamics and Brown-York energy}

The static cavity response follows from the same wall variables \cite{BrownYork1993,BalasubramanianKraus1999}. For
\begin{equation}
ds^2=-f(r)dt^2+\frac{dr^2}{f(r)}+r^2d\phi^2,
\qquad
f(r)=\frac{r^2-r_+^2}{\ell^2},
\label{eq:sec6_staticBTZ}
\end{equation}
the proper wall temperature, quasilocal energy density, pressure, integrated energy, and entropy are
\begin{equation}
T_R=\frac{r_+}{2\pi\ell\sqrt{R^2-r_+^2}},
\qquad
\epsilon=\frac{1}{8\pi G\ell}\left(1-\frac{\sqrt{R^2-r_+^2}}{R}\right),
\qquad
P=\frac{1}{8\pi G\ell}\left(\frac{R}{\sqrt{R^2-r_+^2}}-1\right),
\label{eq:sec6_TR_epsP}
\end{equation}
\begin{equation}
E=\frac{R-\sqrt{R^2-r_+^2}}{4G\ell},
\qquad
S=\frac{\pi r_+}{2G}.
\label{eq:sec6_ES}
\end{equation}
The Euclidean free energy
\begin{equation}
F(R,T_R)=\frac{R}{4G\ell}\left(1-\sqrt{1+4\pi^2\ell^2T_R^2}\right)
\label{eq:sec6_F}
\end{equation}
satisfies the exact Legendre relation
\begin{equation}
F=E-T_RS,
\qquad
S=-\left(\frac{\partial F}{\partial T_R}\right)_R,
\qquad
dE=T_RdS-PdL,
\qquad
L=2\pi R.
\label{eq:sec6_legendre}
\end{equation}
Thus the Brown-York charge is the internal energy of the finite wall system \cite{BrownYork1993,York1986}.

The heat capacity at fixed circumference is
\begin{equation}
C_L=T_R\left(\frac{\partial S}{\partial T_R}\right)_L
=\frac{\pi^2\ell RT_R}{G(1+4\pi^2\ell^2T_R^2)^{3/2}}
=\frac{\pi r_+}{2G}\left(1-\frac{r_+^2}{R^2}\right).
\label{eq:sec6_CL}
\end{equation}
Since \(0<r_+<R\), one has \(C_L>0\). Equivalently,
\begin{equation}
\left(\frac{\partial^2F}{\partial T_R^2}\right)_R
=-\frac{C_L}{T_R}<0,
\label{eq:sec6_Fconcavity}
\end{equation}
which is the canonical stability condition written in wall variables.

The limits are physically transparent. In the massless limit \(r_+\to0\), \(E,P,C_L\to0\). For \(R\gg r_+,\ell\),
\begin{equation}
\epsilon=P=\frac{\pi c}{6}T_R^2+O(R^{-4}),
\qquad
c=\frac{3\ell}{2G},
\label{eq:sec6_UV}
\end{equation}
so the wall theory approaches the undeformed thermal CFT equation of state \cite{BrownHenneaux1986}. For \(r_+\to R^-\), the local temperature and pressure diverge, while
\begin{equation}
E\to\frac{R}{4G\ell},
\qquad
C_L\to0.
\label{eq:sec6_nearwall}
\end{equation}
The static cavity is therefore locally stable throughout the physical interval, with a finite quasilocal energy even in the strongly redshifted near-wall regime.

\section{Rotating BTZ in a cavity and the grand-canonical ensemble}

The rotating cavity is naturally grand canonical because the finite Euclidean wall torus has both a thermal cycle and an angular twist \cite{BanadosTeitelboimZanelli1992,BanadosHenneauxTeitelboimZanelli1993,HuangTao2022}. The control data are \((R,T_R,\Omega_R)\), and the thermodynamic potential is the wall grand potential.

Using the rotating BTZ metric of Section~4, the local intensive variables are
\begin{equation}
T_R=\frac{R(r_+^2-r_-^2)}{2\pi\ell r_+\sqrt{(R^2-r_+^2)(R^2-r_-^2)}},
\qquad
\Omega_R=\frac{r_-\sqrt{R^2-r_+^2}}{r_+R\sqrt{R^2-r_-^2}}.
\label{eq:sec7_TROR}
\end{equation}
The extensive variables are
\begin{equation}
E(R)=\frac{R}{4G\ell}\left(1-\frac{\sqrt{(R^2-r_+^2)(R^2-r_-^2)}}{R^2}\right),
\qquad
J=\frac{r_+r_-}{4G\ell},
\qquad
S=\frac{\pi r_+}{2G}.
\label{eq:sec7_EJS}
\end{equation}
They obey
\begin{equation}
dE=T_RdS+\Omega_RdJ-PdL,
\qquad
L=2\pi R.
\label{eq:sec7_firstlaw}
\end{equation}

To show that the wall data determine a unique smooth rotating saddle, set
\begin{equation}
x=\frac{r_+}{R},
\qquad
y=\frac{r_-}{R},
\qquad
a=2\pi\ell T_R,
\qquad
q=R\Omega_R.
\label{eq:sec7_dimless}
\end{equation}
Then
\begin{equation}
a=\frac{x^2-y^2}{x\sqrt{(1-x^2)(1-y^2)}},
\qquad
q=\frac{y\sqrt{1-x^2}}{x\sqrt{1-y^2}},
\label{eq:sec7_aq}
\end{equation}
and inversion gives
\begin{equation}
r_- =\frac{2\pi\ell R^2T_R\Omega_R}{1-R^2\Omega_R^2},
\qquad
r_+ =\frac{2\pi\ell RT_R}{\sqrt{(1-R^2\Omega_R^2)(1+4\pi^2\ell^2T_R^2-R^2\Omega_R^2)}}.
\label{eq:sec7_inverse}
\end{equation}
The physical wall domain is
\begin{equation}
R^2\Omega_R^2<1,
\qquad
1+4\pi^2\ell^2T_R^2-R^2\Omega_R^2>0.
\label{eq:sec7_wall_domain}
\end{equation}

The grand potential
\begin{equation}
\mathcal G_R=E-T_RS-\Omega_RJ
\label{eq:sec7_GRdef}
\end{equation}
is
\begin{equation}
\mathcal G_R
=\frac{R}{4G\ell}\left(1-\sqrt{\frac{R^2-r_-^2}{R^2-r_+^2}}\right)
=\frac{R}{4G\ell}\left[
1-
\sqrt{\frac{1+4\pi^2\ell^2T_R^2-R^2\Omega_R^2}{1-R^2\Omega_R^2}}
\right].
\label{eq:sec7_GRwall}
\end{equation}
For \(\Omega_R=0\), this reduces to the static Helmholtz free energy. The fixed-boundary partition function has the semiclassical form
\begin{equation}
Z(R,\beta_R,\Omega_R)\sim\exp[-I_E^{\rm on\mbox{-}shell}],
\qquad
I_E^{\rm on\mbox{-}shell}=\beta_R\mathcal G_R,
\label{eq:sec7_Zgrand}
\end{equation}
and
\begin{equation}
S=-\left(\frac{\partial\mathcal G_R}{\partial T_R}\right)_{R,\Omega_R},
\qquad
J=-\left(\frac{\partial\mathcal G_R}{\partial\Omega_R}\right)_{R,T_R}.
\label{eq:sec7_SJfromG}
\end{equation}

For \(R\gg r_+,r_-,\ell\),
\begin{equation}
T_R=\frac{\ell T_H}{R}+O(R^{-3}),
\qquad
\Omega_R=\frac{\ell\Omega_H}{R}+O(R^{-3}),
\qquad
\mathcal G_R=-\frac{\ell}{R}(M-\Omega_HJ)+O(R^{-3}).
\label{eq:sec7_largeR}
\end{equation}
In the extremal approach \(r_-\to r_+\),
\begin{equation}
T_R\to0,
\qquad
\Omega_R\to\frac{1}{R}.
\label{eq:sec7_extreme}
\end{equation}
Thus the rotating BTZ cavity is an exact finite-radius grand-canonical system whose equilibrium data are determined entirely by the wall torus.

\section{Off-shell free-energy landscape and local stability}

The off-shell cavity construction keeps the wall data fixed while varying the horizon parameters. The reduced potentials are finite-dimensional representatives of the Euclidean action on the BTZ family; their stationary points impose smoothness, and their Hessians encode local stability \cite{BrownComerMartinezMelmedWhitingYork1990,HuangTao2022}.

For the static ensemble, the off-shell parameter is \(0<r_+<R\). The reduced Helmholtz potential is
\begin{equation}
F_R^{\,b}(R,T_R;r_+)
=E(R,r_+)-T_RS(r_+)
=\frac{R-\sqrt{R^2-r_+^2}}{4G\ell}-\frac{\pi T_R r_+}{2G}.
\label{eq:sec8_Fb_static}
\end{equation}
The stationary equation is
\begin{equation}
\frac{\partial F_R^{\,b}}{\partial r_+}=0
\quad\Longleftrightarrow\quad
T_R=\frac{r_+}{2\pi\ell\sqrt{R^2-r_+^2}},
\label{eq:sec8_static_saddle}
\end{equation}
which is exactly the smooth Euclidean BTZ condition. Moreover,
\begin{equation}
\frac{\partial^2F_R^{\,b}}{\partial r_+^2}
=
\frac{R^2}{4G\ell(R^2-r_+^2)^{3/2}}>0,
\label{eq:sec8_d2F}
\end{equation}
so the smooth static saddle is the unique local minimum. At equilibrium,
\begin{equation}
\left.\frac{\partial^2F_R^{\,b}}{\partial r_+^2}\right|_\ast
=\frac{\pi^2T_R}{4G^2C_L},
\qquad
C_L=\frac{\pi r_+^\ast}{2G}\left(1-\frac{(r_+^\ast)^2}{R^2}\right),
\label{eq:sec8_curvature_CL}
\end{equation}
so convexity and positive heat capacity are the same local statement in different variables.

For the rotating ensemble, the off-shell parameters obey \(0\le r_-<r_+<R\), and
\begin{equation}
G_R^{\,b}(R,T_R,\Omega_R;r_+,r_-)
=
\frac{R}{4G\ell}\left(1-\frac{\Delta_R}{R^2}\right)
-\frac{\pi T_Rr_+}{2G}
-\frac{\Omega_Rr_+r_-}{4G\ell},
\label{eq:sec8_Gb}
\end{equation}
where \(\Delta_R=\sqrt{(R^2-r_+^2)(R^2-r_-^2)}\). Differentiation gives
\begin{equation}
\frac{\partial G_R^{\,b}}{\partial r_+}
=
\frac{r_+(R^2-r_-^2)}{4G\ell R\Delta_R}
-\frac{\pi T_R}{2G}
-\frac{\Omega_Rr_-}{4G\ell},
\label{eq:sec8_dGdrplus}
\end{equation}
\begin{equation}
\frac{\partial G_R^{\,b}}{\partial r_-}
=
\frac{r_-(R^2-r_+^2)}{4G\ell R\Delta_R}
-\frac{\Omega_Rr_+}{4G\ell}.
\label{eq:sec8_dGdrminus}
\end{equation}
The stationary equations reproduce the wall smoothness conditions,
\begin{equation}
T_R=\frac{R(r_+^2-r_-^2)}{2\pi\ell r_+\sqrt{(R^2-r_+^2)(R^2-r_-^2)}},
\qquad
\Omega_R=\frac{r_-\sqrt{R^2-r_+^2}}{r_+R\sqrt{R^2-r_-^2}}.
\label{eq:sec8_rot_saddle}
\end{equation}
The Hessian at the stationary point has
\begin{equation}
H_{++}=\frac{R\sqrt{R^2-(r_-^\ast)^2}}{4G\ell(R^2-(r_+^\ast)^2)^{3/2}}>0,
\label{eq:sec8_Hpp}
\end{equation}
and
\begin{equation}
\det H
=
\frac{R^2\bigl((r_+^\ast)^2-(r_-^\ast)^2\bigr)}
{16G^2\ell^2(r_+^\ast)^2(R^2-(r_+^\ast)^2)(R^2-(r_-^\ast)^2)}.
\label{eq:sec8_detH}
\end{equation}
Both quantities are positive throughout \(0\le r_-^\ast<r_+^\ast<R\), so the rotating grand potential has a strict local minimum over the non-extremal physical domain.

The limiting behavior matches the expected geometry. At low static temperature, \(r_+^\ast=2\pi\ell RT_R+O(T_R^3)\) and the minimum lies deep inside the cavity. At high temperature, \(r_+^\ast=R[1-(8\pi^2\ell^2T_R^2)^{-1}+O(T_R^{-4})]\), and the saddle is pinned close to the wall. Near extremality, with \(r_-^\ast=r_+^\ast-\varepsilon\), the determinant \eqref{eq:sec8_detH} is linear in the extremality gap, so one local fluctuation direction softens as the zero-temperature rotating state is approached. The off-shell construction therefore provides the local variational backbone for the finite-cutoff thermodynamic ensemble; the next section turns to the global competition between inequivalent smooth fillings.

\section{Finite-size Hawking-Page transition}

The off-shell analysis of the previous section establishes that the static BTZ branch is locally stable everywhere inside the cavity. The remaining question is therefore global rather than local: among the two smooth Euclidean fillings of the same wall torus, namely Euclidean BTZ and thermal AdS\(_3\), which one has lower free energy at fixed wall data? The answer defines the finite-radius analogue of the Hawking-Page transition and gives the sharpest global consequence of the cavity formulation. In the present setting the transition is controlled directly by the wall torus itself, so the critical scale is set by the proper circumference of the cavity rather than by asymptotic data \cite{WhitingYork1988,HawkingPage1983,Witten1998Thermal,HuangTao2022,MaloneyWitten2010}.

The purpose of this section is to determine that phase boundary exactly and to characterize the resulting transition in quasilocal thermodynamic variables. Starting from the wall free energies already derived in the Euclidean analysis, we solve the equality condition, obtain the critical temperature and critical horizon radius in closed form, and compute the entropy jump and latent heat. The resulting transition is first order, globally determined, and universal when written in wall units.

For the static cavity at fixed wall radius \(R\), the relevant smooth Euclidean saddles are Euclidean BTZ and thermal AdS\(_3\). Their wall free energies are
\begin{equation}
F_{\mathrm{BH}}(R,T_{R})
=
\frac{R}{4G\ell}
\left(
1-\sqrt{1+4\pi^{2}\ell^{2}T_{R}^{2}}
\right),
\label{eq:sec9_FBH}
\end{equation}
\begin{equation}
F_{\mathrm{AdS}}(R)
=
\frac{1}{4G\ell}
\left(
R-\sqrt{R^{2}+\ell^{2}}
\right).
\label{eq:sec9_FAdS}
\end{equation}
The difference is therefore
\begin{equation}
\Delta F(R,T_{R})
:=
F_{\mathrm{BH}}(R,T_{R})-F_{\mathrm{AdS}}(R)
=
\frac{1}{4G\ell}
\left(
\sqrt{R^{2}+\ell^{2}}
-
R\sqrt{1+4\pi^{2}\ell^{2}T_{R}^{2}}
\right).
\label{eq:sec9_DeltaF}
\end{equation}
Since the static BTZ branch is locally stable for every \(0<r_{+}<R\), the phase boundary is determined solely by the sign change of \(\Delta F\). The transition condition \(\Delta F=0\) gives
\begin{equation}
R\sqrt{1+4\pi^{2}\ell^{2}T_{c}^{2}}
=
\sqrt{R^{2}+\ell^{2}},
\label{eq:sec9_transitioneq}
\end{equation}
hence
\begin{equation}
T_{c}(R)=\frac{1}{2\pi R}.
\label{eq:sec9_Tc}
\end{equation}
This is the central result of the section. The critical temperature depends only on the proper wall size and is independent of the AdS radius when expressed in locally measured wall variables.

The corresponding critical black hole is obtained from the static wall relation
\begin{equation}
T_{R}
=
\frac{r_{+}}{2\pi \ell\sqrt{R^{2}-r_{+}^{2}}},
\label{eq:sec9_TRrplus}
\end{equation}
which at \(T_{R}=T_{c}\) yields
\begin{equation}
r_{+,c}(R)
=
\frac{\ell R}{\sqrt{R^{2}+\ell^{2}}}.
\label{eq:sec9_rplusc}
\end{equation}
Equations \eqref{eq:sec9_Tc} and \eqref{eq:sec9_rplusc} identify the precise point at which the dominant smooth Euclidean filling changes. For temperatures below the critical value,
\begin{equation}
T_{R}<T_{c}(R)\quad \Longrightarrow \quad F_{\mathrm{AdS}}<F_{\mathrm{BH}},
\label{eq:sec9_lowTdom}
\end{equation}
whereas for temperatures above it,
\begin{equation}
T_{R}>T_{c}(R)\quad \Longrightarrow \quad F_{\mathrm{BH}}<F_{\mathrm{AdS}}.
\label{eq:sec9_highTdom}
\end{equation}
The transition is therefore global in origin. Every BTZ saddle is locally stable, though thermal AdS\(_3\) dominates the ensemble at sufficiently low wall temperature.

The geometric meaning of \eqref{eq:sec9_Tc} is immediate. The wall torus has proper thermal circumference \(\beta_{R}=T_{R}^{-1}\) and spatial circumference \(L=2\pi R\). At the transition,
\begin{equation}
\beta_{R}=L.
\label{eq:sec9_betaeqL}
\end{equation}
The phase boundary is therefore the point at which the two cycles of the same wall torus have equal proper length. The finite-size Hawking-Page transition is thus the modular crossover of the wall torus, written in quasilocal thermodynamic language \cite{HawkingPage1983,Witten1998Thermal,MaloneyWitten2010}. The dominance changes because the two possible smooth fillings of the same torus exchange which cycle is contractible in the preferred saddle.

The transition is first order. Thermal AdS\(_3\) carries no entropy in this ensemble because its free energy is independent of \(T_{R}\), while the BTZ branch carries the Bekenstein--Hawking entropy
\begin{equation}
S_{\mathrm{BH}}(R,T_{R})
=
-\left(\frac{\partial F_{\mathrm{BH}}}{\partial T_{R}}\right)_{R}
=
\frac{\pi r_{+}}{2G}.
\label{eq:sec9_SBH}
\end{equation}
At the critical point this becomes
\begin{equation}
\Delta S
=
S_{\mathrm{BH}}(R,T_{c})
=
\frac{\pi \ell R}{2G\sqrt{R^{2}+\ell^{2}}}.
\label{eq:sec9_DeltaS}
\end{equation}
The latent heat is therefore
\begin{equation}
Q_{\mathrm{latent}}
=
T_{c}\,\Delta S
=
\frac{\ell}{4G\sqrt{R^{2}+\ell^{2}}}.
\label{eq:sec9_Qlatent}
\end{equation}
Equivalently, since the free energies coincide at the phase boundary, the same quantity is the jump in quasilocal internal energy,
\begin{equation}
Q_{\mathrm{latent}}
=
E_{\mathrm{BH}}(R,r_{+,c})-E_{\mathrm{AdS}}(R),
\label{eq:sec9_QlatentE}
\end{equation}
with
\begin{equation}
E_{\mathrm{BH}}(R,r_{+})
=
\frac{R-\sqrt{R^{2}-r_{+}^{2}}}{4G\ell},
\qquad
E_{\mathrm{AdS}}(R)=F_{\mathrm{AdS}}(R).
\label{eq:sec9_Esaddles}
\end{equation}
The phase transition is therefore first order in the standard thermodynamic sense.

Several limiting regimes clarify the result. In the low-temperature limit \(T_{R}\to0^{+}\),
\begin{equation}
F_{\mathrm{BH}}(R,T_{R})\to0,
\qquad
F_{\mathrm{AdS}}(R)=\frac{1}{4G\ell}\left(R-\sqrt{R^{2}+\ell^{2}}\right)<0,
\label{eq:sec9_lowT}
\end{equation}
so thermal AdS\(_3\) dominates. In the high-temperature regime \(2\pi \ell T_{R}\gg1\),
\begin{equation}
F_{\mathrm{BH}}(R,T_{R})
=
-\frac{\pi R}{2G}\,T_{R}
+\frac{R}{4G\ell}
+O(T_{R}^{-1}),
\label{eq:sec9_highT}
\end{equation}
and the BTZ saddle dominates. The asymptotic AdS limit is also informative. For \(R\gg \ell\),
\begin{equation}
r_{+,c}(R)
=
\ell\left(
1-\frac{\ell^{2}}{2R^{2}}+O(R^{-4})
\right),
\label{eq:sec9_largeR_rplusc}
\end{equation}
and since
\begin{equation}
T_{R}=\frac{r_{+}}{2\pi \ell R}+O(R^{-3}),
\label{eq:sec9_largeR_TR}
\end{equation}
the critical condition \(T_{R}=1/(2\pi R)\) becomes \(r_{+}=\ell+O(R^{-2})\). The finite-radius phase boundary therefore reduces smoothly to the standard asymptotic Hawking--Page threshold. In the opposite regime \(R\ll \ell\),
\begin{equation}
r_{+,c}(R)
=
R\left(
1-\frac{R^{2}}{2\ell^{2}}+O(R^{4}\ell^{-4})
\right),
\label{eq:sec9_smallR_rplusc}
\end{equation}
so the critical black hole sits close to the wall and the transition probes a highly local portion of the geometry. In that regime the critical temperature
\begin{equation}
T_{c}(R)=\frac{1}{2\pi R}
\label{eq:sec9_smallR_Tc}
\end{equation}
is much larger than the AdS scale, and the phase structure is controlled almost entirely by the wall torus itself.

\begin{figure}[t]
  \centering
  \includegraphics[width=0.78\linewidth]{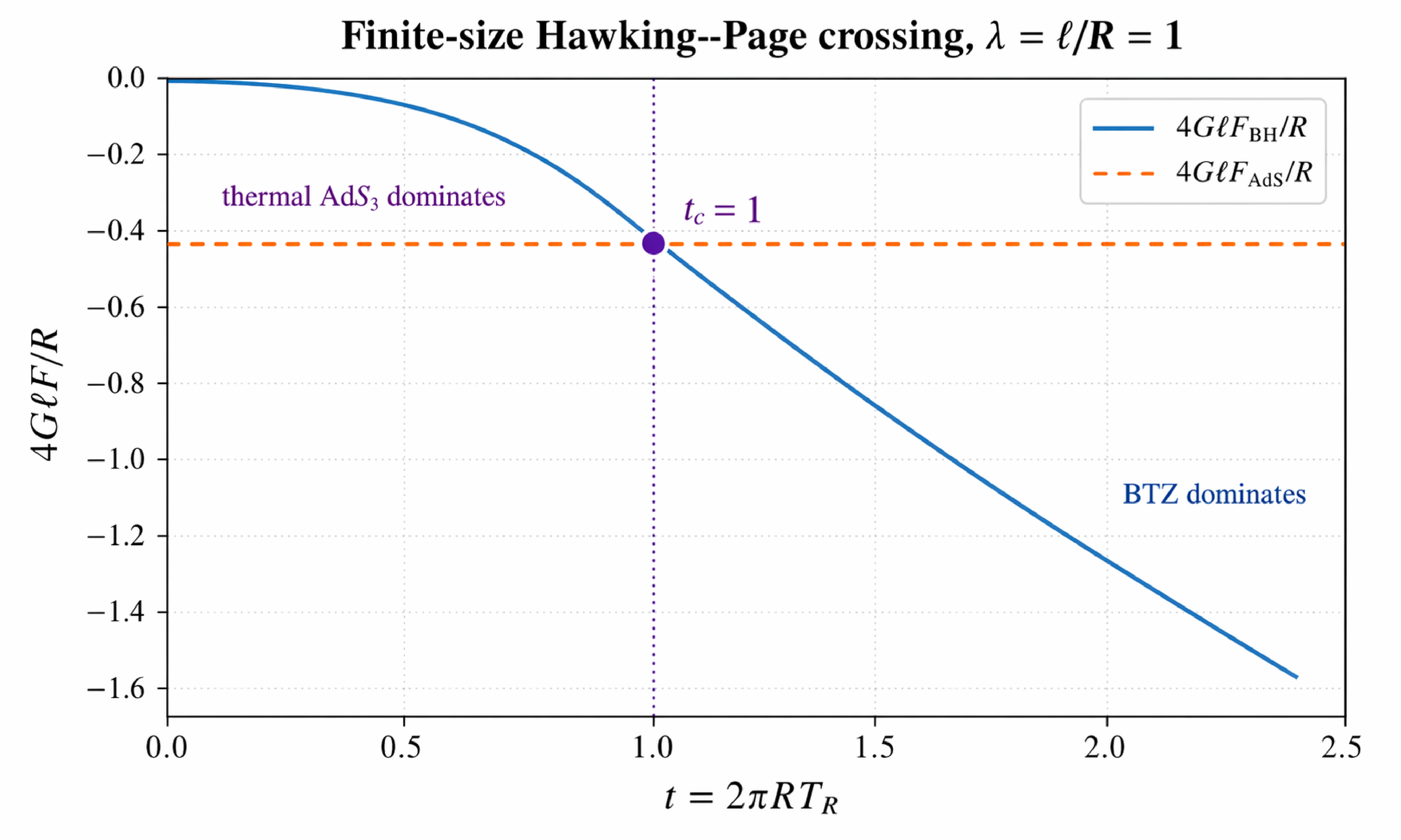}
  \caption{
  Dimensionless free energies of the static finite-radius ensemble in wall units,
  shown for \(\lambda=\ell/R=1\). The BTZ branch obeys
  \(4G\ell F_{\rm BH}/R=1-\sqrt{1+\lambda^{2}t^{2}}\), while thermal AdS\(_3\)
  obeys \(4G\ell F_{\rm AdS}/R=1-\sqrt{1+\lambda^{2}}\), with
  \(t=2\pi R T_R\). The crossing occurs at \(t_c=1\), equivalently
  \(T_c(R)=1/(2\pi R)\), where the proper thermal and spatial cycles of the wall
  torus have equal length. Thermal AdS\(_3\) dominates for \(t<1\), while the
  BTZ saddle dominates for \(t>1\).
  }
  \label{fig:finite-size-hawking-page}
\end{figure}

A compact universal parametrization is obtained by introducing the dimensionless wall variables
\begin{equation}
t:=2\pi R T_{R},
\qquad
\lambda:=\frac{\ell}{R}.
\label{eq:sec9_tlambda}
\end{equation}
Then the free energies read
\begin{equation}
\frac{4G\ell}{R}F_{\mathrm{BH}}
=
1-\sqrt{1+\lambda^{2}t^{2}},
\qquad
\frac{4G\ell}{R}F_{\mathrm{AdS}}
=
1-\sqrt{1+\lambda^{2}},
\label{eq:sec9_wallunits}
\end{equation}
and the phase boundary is simply
\begin{equation}
t_{c}=1.
\label{eq:sec9_tc}
\end{equation}
Once all quantities are expressed in wall units, the transition occurs at unit temperature of the boundary torus. This is the sharpest manifestation of universality in the finite-radius formulation. The free-energy crossing is shown in Figure~\ref{fig:finite-size-hawking-page}; in wall units the transition occurs at the universal value \(t_c=1\).

The finite-size Hawking-Page transition therefore has a very clean interpretation. A single wall torus admits two smooth bulk realizations, Euclidean BTZ and thermal AdS\(_3\), and the dominant realization changes exactly when the proper thermal cycle equals the proper spatial cycle. The transition is first order, the BTZ branch remains locally stable across the entire physical domain, and the critical temperature is fixed by the inverse circumference of the cavity. This global phase structure will be reinterpreted in the next sections as the phase structure of the finite-cutoff dual theory, with the wall radius playing the role of the scale and size of that system simultaneously.

\section{Holographic RG interpretation}

The cavity radius enters the finite-cutoff description as more than a geometric parameter. At fixed bulk state, changing the location of the wall changes the local normalization of time, the local chemical potentials, and the Brown-York stress tensor measured on the wall, while leaving the underlying BTZ solution unchanged. The radial position therefore acts as a genuine running scale for the finite system defined on the wall. In three-dimensional Einstein gravity this statement is especially sharp because there are no local propagating bulk gravitons, so radial evolution closes on the boundary stress tensor and its thermodynamic observables. The holographic renormalization-group interpretation is therefore not an additional conjectural layer placed on top of the cavity thermodynamics; it is the direct reading of the finite-radius wall dynamics in scale-dependent language \cite{deHaroSkenderisSolodukhin2001,Skenderis2002,BalasubramanianKraus1999,McGoughMezeiVerlinde2018,Taylor2018,Shyam2017}.

The aim of the present section is to formulate that running explicitly in wall variables and to isolate the corresponding beta functions. The central observation is that once the bulk state is fixed by \((r_{+},r_{-})\), the radial motion of the wall generates exact first-order flow equations for the quasilocal thermodynamic data. These equations organize the finite-cutoff dynamics of the wall theory and show how the ultraviolet conformal regime is recovered at large radius while the finite-cutoff deformation grows as the wall is moved inward.

For the rotating BTZ family, the local wall quantities are
\begin{equation}
\epsilon(R)
=
\frac{1}{8\pi G\ell}
\left(
1-\frac{\Delta_{R}}{R^{2}}
\right),
\qquad
j(R)=\frac{r_{+}r_{-}}{8\pi G\ell R^{2}},
\qquad
p(R)
=
\frac{1}{8\pi G\ell}
\left(
\frac{R^{4}-r_{+}^{2}r_{-}^{2}}{R^{2}\Delta_{R}}-1
\right),
\label{eq:sec10_epsjp}
\end{equation}
with
\begin{equation}
\Delta_{R}:=\sqrt{(R^{2}-r_{+}^{2})(R^{2}-r_{-}^{2})},
\label{eq:sec10_DeltaR}
\end{equation}
and the corresponding integrated observables are
\begin{equation}
E(R)=2\pi R\,\epsilon(R)=\frac{R}{4G\ell}\left(1-\frac{\Delta_{R}}{R^{2}}\right),
\qquad
J=2\pi R^{2}j(R)=\frac{r_{+}r_{-}}{4G\ell}.
\label{eq:sec10_EJ}
\end{equation}
The proper wall temperature and angular potential are
\begin{equation}
T_{R}
=
\frac{R(r_{+}^{2}-r_{-}^{2})}{2\pi \ell r_{+}\Delta_{R}},
\qquad
\Omega_{R}
=
\frac{r_{-}\sqrt{R^{2}-r_{+}^{2}}}{r_{+}R\sqrt{R^{2}-r_{-}^{2}}}.
\label{eq:sec10_TROR}
\end{equation}
These quantities are defined intrinsically at the wall and are therefore the natural running observables of the finite-cutoff system.

The first flow equation follows directly from the finite-radius first law
\begin{equation}
dE=T_{R}\,dS+\Omega_{R}\,dJ-P\,dL,
\qquad
L=2\pi R,
\label{eq:sec10_firstlaw}
\end{equation}
when the bulk state is held fixed. Since \(S\) and \(J\) are then constant, one finds
\begin{equation}
\left(\frac{\partial E}{\partial R}\right)_{r_{+},r_{-}}
=
-2\pi P.
\label{eq:sec10_dEdR}
\end{equation}
Using \(E(R)=2\pi R\,\epsilon(R)\), this gives the local energy-density flow
\begin{equation}
R\left(\frac{\partial \epsilon}{\partial R}\right)_{r_{+},r_{-}}
=
-(\epsilon+p).
\label{eq:sec10_epsflow}
\end{equation}
For the momentum density, the conservation of the integrated angular momentum \(J=2\pi R^{2}j(R)\) implies
\begin{equation}
R\left(\frac{\partial j}{\partial R}\right)_{r_{+},r_{-}}
=
-2j.
\label{eq:sec10_jflow}
\end{equation}
Equations \eqref{eq:sec10_epsflow} and \eqref{eq:sec10_jflow} are the finite-cutoff flow equations for the extensive and momentum sectors of the homogeneous wall state.

The intensive variables run as well. Differentiating \eqref{eq:sec10_TROR} at fixed \((r_{+},r_{-})\) yields
\begin{equation}
R\left(\frac{\partial T_{R}}{\partial R}\right)_{r_{+},r_{-}}
=
-
T_{R}\,
\frac{R^{4}-r_{+}^{2}r_{-}^{2}}
{(R^{2}-r_{+}^{2})(R^{2}-r_{-}^{2})},
\label{eq:sec10_TRflow}
\end{equation}
and
\begin{equation}
R\left(\frac{\partial \Omega_{R}}{\partial R}\right)_{r_{+},r_{-}}
=
\Omega_{R}
\left[
\frac{R^{2}(r_{+}^{2}-r_{-}^{2})}
{(R^{2}-r_{+}^{2})(R^{2}-r_{-}^{2})}
-1
\right].
\label{eq:sec10_ORflow}
\end{equation}
The temperature flow is always negative in the physical domain \(0\le r_{-}\le r_{+}<R\), so the local wall temperature decreases monotonically as the wall is moved outward. This is the Tolman redshift written as a beta function. The angular-potential flow changes sign according to the location in parameter space, though it is likewise determined exactly by the geometry of the fixed bulk state.

The same running is visible in the Hamilton-Jacobi form of the wall dynamics. The universal finite-cutoff constraint derived in Section \ref{sec:section_3}  reads
\begin{equation}
T^{i}{}_{i}
=
\frac{\ell}{16\pi G}R[h]
+
4\pi G\ell
\left(
T_{ij}T^{ij}-\bigl(T^{i}{}_{i}\bigr)^{2}
\right).
\label{eq:sec10_traceHJ}
\end{equation}
On the flat wall cylinder of the BTZ cavity, \(R[h]=0\), and for homogeneous states this becomes
\begin{equation}
p-\epsilon
=
8\pi G\ell\left(\epsilon p-j^{2}\right).
\label{eq:sec10_traceflat}
\end{equation}
Equation \eqref{eq:sec10_traceflat} is the local finite-cutoff trace law that controls the departure from conformality along the radial flow \cite{McGoughMezeiVerlinde2018,Taylor2018,Shyam2017}. It is the exact relation that closes the running of the homogeneous wall state. In the ultraviolet limit \(R\gg \ell,r_{+},r_{-}\), the wall observables behave as
\begin{equation}
\epsilon(R)
=
\frac{r_{+}^{2}+r_{-}^{2}}{16\pi G\ell R^{2}}
+O(R^{-4}),
\qquad
j(R)
=
\frac{r_{+}r_{-}}{8\pi G\ell R^{2}},
\qquad
p(R)
=
\frac{r_{+}^{2}+r_{-}^{2}}{16\pi G\ell R^{2}}
+O(R^{-4}),
\label{eq:sec10_UVstress}
\end{equation}
so that
\begin{equation}
p-\epsilon=O(R^{-4}).
\label{eq:sec10_tracelessUV}
\end{equation}
The trace therefore vanishes asymptotically, and the wall theory approaches the ultraviolet conformal equation of state. Using the Brown-Henneaux central charge
\begin{equation}
c=\frac{3\ell}{2G},
\label{eq:sec10_c}
\end{equation}
the static sector reduces to
\begin{equation}
\epsilon=p=\frac{\pi c}{6}\,T_{R}^{2}+O(R^{-4}),
\label{eq:sec10_CFTeq}
\end{equation}
which is the thermal stress tensor of the undeformed CFT measured in wall variables \cite{BrownHenneaux1986,Cardy1986}.

The inward flow away from this ultraviolet regime is equally transparent. As \(R\downarrow r_{+}\), one has
\begin{equation}
\Delta_{R}
=
\sqrt{2r_{+}(r_{+}^{2}-r_{-}^{2})(R-r_{+})}
+O\!\left((R-r_{+})^{3/2}\right),
\label{eq:sec10_nearwallDelta}
\end{equation}
and therefore
\begin{equation}
T_{R}
=
\frac{\sqrt{r_{+}^{2}-r_{-}^{2}}}{2\pi\ell\sqrt{2r_{+}(R-r_{+})}}
+O\!\left((R-r_{+})^{1/2}\right),
\label{eq:sec10_nearwallT}
\end{equation}
\begin{equation}
p(R)
=
\frac{\sqrt{r_{+}^{2}-r_{-}^{2}}}{8\pi G\ell\sqrt{2r_{+}(R-r_{+})}}
+O\!\left((R-r_{+})^{1/2}\right),
\label{eq:sec10_nearwallp}
\end{equation}
while
\begin{equation}
E(R)\longrightarrow \frac{r_{+}}{4G\ell}
\qquad
(R\downarrow r_{+}).
\label{eq:sec10_nearwallE}
\end{equation}
The local intensive quantities diverge while the integrated energy remains finite. The radial flow therefore strengthens the deformation of the wall theory as the cutoff is moved inward, though the underlying bulk geometry remains smooth and the wall system stays finite.

\begin{figure}[t]
  \centering
  \includegraphics[width=0.78\linewidth]{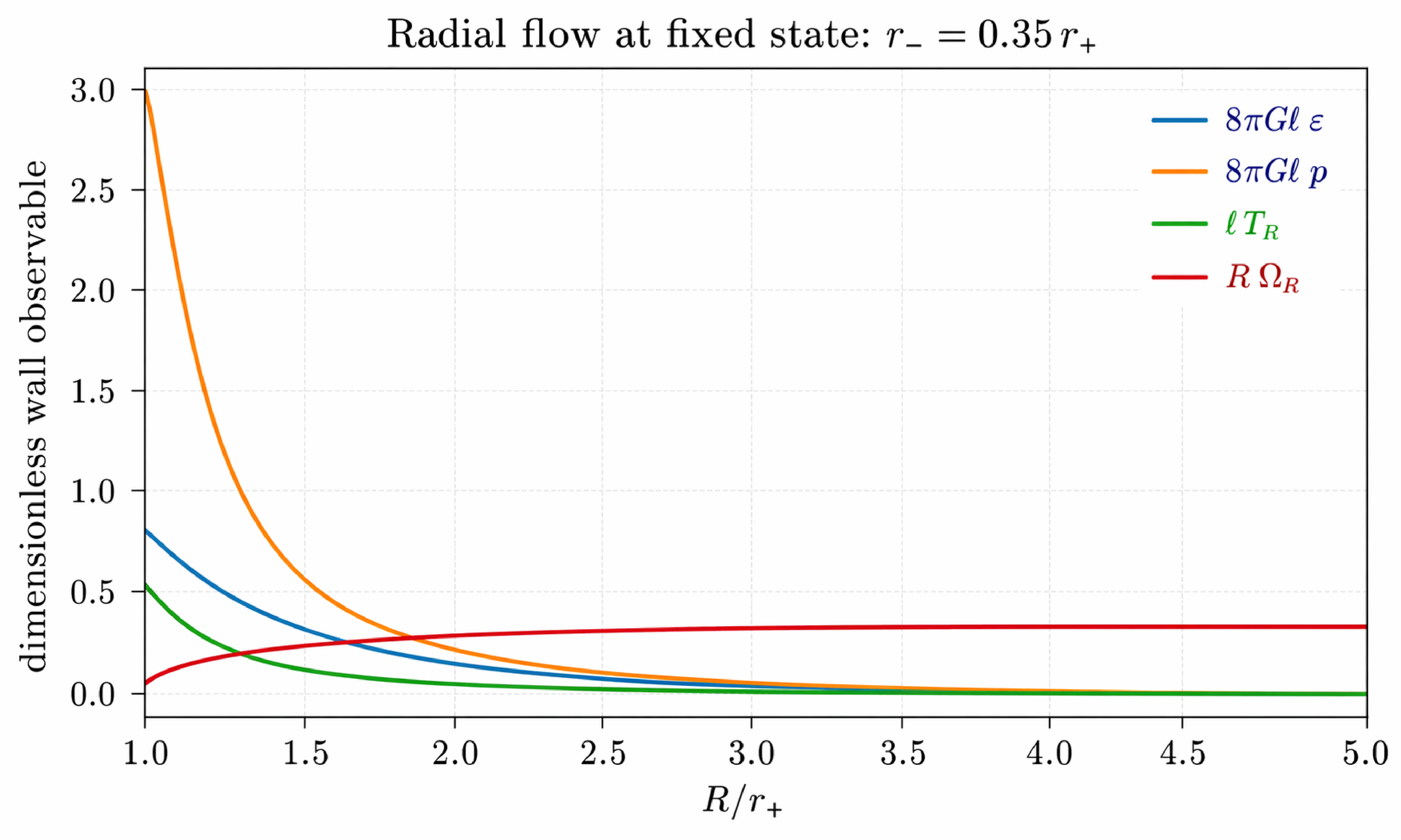}
  \caption{
  Radial flow of dimensionless wall observables at fixed BTZ state, illustrated
  for \(r_-=0.35\,r_+\). The plotted quantities are
  \(8\pi G\ell\,\epsilon\), \(8\pi G\ell\,p\), \(\ell T_R\), and \(R\Omega_R\)
  as functions of \(R/r_+\). As the wall is moved outward, the local stress tensor
  approaches the ultraviolet conformal regime. As the wall approaches the outer
  horizon, the pressure and local temperature increase by the Tolman blueshift,
  while the integrated Brown--York energy remains finite.
  }
  \label{fig:radial-flow-wall-observables}
\end{figure}

The same running controls the finite-size Hawking-Page scale. Since
\begin{equation}
T_{c}(R)=\frac{1}{2\pi R},
\label{eq:sec10_Tc}
\end{equation}
one has
\begin{equation}
R\frac{dT_{c}}{dR}=-T_{c}.
\label{eq:sec10_Tcflow}
\end{equation}
The phase boundary itself therefore scales homogeneously with the wall position, as expected for a finite thermodynamic system whose physical size is \(L=2\pi R\). In this sense the radial coordinate governs not only the running of local observables within a fixed phase but also the running of the global transition scale of the finite-cutoff system. Figure~\ref{fig:radial-flow-wall-observables} illustrates the same radial running in dimensionless wall variables for a representative rotating state.

The holographic RG interpretation may now be stated succinctly. The bulk parameters \((r_{+},r_{-})\) label the state, while the wall position \(R\) determines the scale at which the stress tensor and thermodynamic observables are measured. The Brown-York tensor is conserved, its trace is fixed by the Hamilton-Jacobi identity, and the running of \((\epsilon,j,T_{R},\Omega_{R})\) is governed by exact first-order flow equations. At large \(R\), the trace disappears and the wall theory approaches the ultraviolet CFT. As the wall moves inward, the quadratic composite \(\epsilon p-j^{2}\) becomes important and the system flows away from conformality. The cavity therefore realizes holographic RG flow in a form that is completely explicit and entirely thermodynamic: the running observables are the quasilocal energy density, momentum density, temperature, pressure, and angular potential measured on the wall.

This section fixes the scale-dependent interpretation of the finite-radius construction. The next section sharpens that interpretation by identifying the same wall stress tensor with the exact stress tensor of the finite-cutoff dual theory and by making the relation to the \(T\bar T\) deformation explicit.

\section{Finite-cutoff AdS\(_3\)/CFT\(_2\) and \(T\bar T\)-deformed interpretation}

At finite radius, the cavity acquires a boundary interpretation that is especially rigid in AdS\(_3\). The wall \(\Sigma_{R}\) is simultaneously a physical thermodynamic boundary and the locus at which the dual stress tensor is defined. In three-dimensional Einstein gravity this identification is unusually sharp because radial evolution is governed entirely by the constraints, so the finite-cutoff dynamics closes directly on the renormalized Brown-York tensor \cite{BalasubramanianKraus1999,deHaroSkenderisSolodukhin2001,Skenderis2002,McGoughMezeiVerlinde2018,Taylor2018,Shyam2017}. The purpose of the present section is to make that identification explicit in the language of finite-volume deformations: the wall theory is placed on the finite cylinder defined by the cavity, its stress tensor obeys the exact quadratic trace-flow law characteristic of \(T\bar T\)-type dynamics, and its energy spectrum takes the standard square-root form with a deformation parameter fixed entirely by the bulk couplings \cite{Zamolodchikov2004,CavagliaNegroSzecsenyiTateo2016,McGoughMezeiVerlinde2018,Shyam2017}.

The point of view adopted here is deliberately narrower than that of the next section. We isolate the exact identification of the wall stress tensor with the stress tensor of the cutoff theory, derive the finite-cutoff spectral map, and identify the gravitational value of the deformation parameter. The Legendre structure of the thermodynamic potentials and the deformed Cardy entropy are then deferred to Section~\ref{sec:sec_12}, where the full finite-cutoff dictionary is written in closed form.

We work with the renormalized Dirichlet action
\begin{equation}
S_{\mathrm{ren}}[g;R]
=
\frac{1}{16\pi G}\int_{M} d^{3}x\,\sqrt{-g}\left(\mathcal{R}[g]+\frac{2}{\ell^{2}}\right)
-\frac{1}{8\pi G}\int_{\Sigma_{R}} d^{2}x\,\sqrt{-h}\,K
-\frac{1}{8\pi G\ell}\int_{\Sigma_{R}} d^{2}x\,\sqrt{-h},
\label{eq:sec11_Sren}
\end{equation}
whose variation at fixed induced metric defines the renormalized Brown--York tensor
\begin{equation}
T^{(R)}_{ij}
=
-\frac{2}{\sqrt{-h}}\frac{\delta S_{\mathrm{ren}}}{\delta h^{ij}}
=
\frac{1}{8\pi G}
\left(
Kh_{ij}-K_{ij}-\frac{1}{\ell}h_{ij}
\right).
\label{eq:sec11_BY}
\end{equation}
For the stationary BTZ family, it is natural to pass to locally corotating proper coordinates on the wall,
\begin{equation}
u=N(R)t,
\qquad
x=R\varphi,
\qquad
x\sim x+L,
\qquad
L=2\pi R,
\label{eq:sec11_ux}
\end{equation}
for which the induced wall metric is the flat cylinder
\begin{equation}
\gamma_{ij}dx^{i}dx^{j}
=
-du^{2}+dx^{2},
\qquad
R[\gamma]=0.
\label{eq:sec11_wallmetric}
\end{equation}
Homogeneity then forces the wall stress tensor to take the form
\begin{equation}
T^{(R)}_{ij}
=
\begin{pmatrix}
\epsilon & -j\\
-j & p
\end{pmatrix}_{(u,x)},
\label{eq:sec11_Twall}
\end{equation}
with integrated wall energy and momentum
\begin{equation}
E=L\,\epsilon,
\qquad
\Pi=L\,j=\frac{J}{R}=\frac{2\pi J}{L}.
\label{eq:sec11_EPi}
\end{equation}
In the boundary interpretation, \(T^{(R)}_{ij}\) is the exact stress tensor of the finite-cutoff theory defined on the cylinder of circumference \(L\).

The universal Hamilton--Jacobi relation derived earlier becomes particularly simple on the flat wall cylinder. Since \(R[\gamma]=0\), one has
\begin{equation}
T^{i}{}_{i}
=
4\pi G\ell
\left(
T_{ij}T^{ij}-\bigl(T^{i}{}_{i}\bigr)^{2}
\right),
\label{eq:sec11_traceHJ}
\end{equation}
which for the homogeneous wall state gives
\begin{equation}
p-\epsilon
=
8\pi G\ell\left(\epsilon p-j^{2}\right).
\label{eq:sec11_traceflow}
\end{equation}
This is exactly the local quadratic flow law expected for a \(T\bar T\)-type deformation \cite{Zamolodchikov2004,CavagliaNegroSzecsenyiTateo2016,McGoughMezeiVerlinde2018,Shyam2017}. It is therefore natural to identify the gravitational deformation parameter as
\begin{equation}
\mu_{\mathrm{grav}}:=8\pi G\ell.
\label{eq:sec11_mugrav}
\end{equation}
The significance of \eqref{eq:sec11_mugrav} is that the deformation strength is fixed by the bulk couplings and requires no additional phenomenological input. The finite-cutoff theory is not merely analogous to a solvable irrelevant deformation; its local stress tensor relation is the exact gravitational realization of that deformation on the wall.

To make the spectral interpretation precise, introduce the undeformed ultraviolet cylinder energy and momentum
\begin{equation}
E_{0}(L):=\frac{2\pi\ell M}{L}=\frac{\ell M}{R},
\qquad
\Pi_{0}(L):=\frac{2\pi J}{L}=\frac{J}{R},
\qquad
E_{0}^{\pm}:=\frac{1}{2}\left(E_{0}\pm\Pi_{0}\right),
\label{eq:sec11_E0Pi0}
\end{equation}
together with the Brown--Henneaux central charge
\begin{equation}
c=\frac{3\ell}{2G}.
\label{eq:sec11_c}
\end{equation}
For the rotating BTZ family,
\begin{equation}
M=\frac{r_{+}^{2}+r_{-}^{2}}{8G\ell^{2}},
\qquad
J=\frac{r_{+}r_{-}}{4G\ell},
\label{eq:sec11_MJ}
\end{equation}
while the quasilocal wall energy is
\begin{equation}
E(R)
=
\frac{R}{4G\ell}\left(1-\frac{\Delta_{R}}{R^{2}}\right),
\qquad
\Delta_{R}^{2}
=
(R^{2}-r_{+}^{2})(R^{2}-r_{-}^{2})
=
R^{4}-8G\ell^{2}MR^{2}+16G^{2}\ell^{2}J^{2}.
\label{eq:sec11_ER}
\end{equation}
Since \(\Pi=J/R=\Pi_{0}\), momentum is protected by spatial translation invariance along the wall cylinder and the deformation acts only on the energy assignment. Eliminating \(M\) and \(J\) in favor of \((E_{0},\Pi_{0})\), and using \(L=2\pi R\) together with \eqref{eq:sec11_mugrav}, one obtains
\begin{equation}
\left(
1-\frac{\mu_{\mathrm{grav}}}{L}E
\right)^{2}
=
1-\frac{2\mu_{\mathrm{grav}}}{L}E_{0}
+\frac{\mu_{\mathrm{grav}}^{2}}{L^{2}}\Pi_{0}^{2}.
\label{eq:sec11_quadratic}
\end{equation}
Choosing the branch that is continuous with the ultraviolet limit \(\mu_{\mathrm{grav}}/L\to0\), the exact finite-cutoff spectrum is
\begin{equation}
E(L;\mu_{\mathrm{grav}},E_{0},\Pi_{0})
=
\frac{L}{\mu_{\mathrm{grav}}}
\left[
1-
\sqrt{
1-\frac{2\mu_{\mathrm{grav}}}{L}E_{0}
+\frac{\mu_{\mathrm{grav}}^{2}}{L^{2}}\Pi_{0}^{2}
}
\right],
\qquad
\Pi=\Pi_{0}.
\label{eq:sec11_spectrum}
\end{equation}
Equation \eqref{eq:sec11_spectrum} is the gravitational square-root spectrum written directly in wall variables. It is the spectral statement underlying the thermodynamic identification of the cavity with a finite-cutoff two-dimensional system \cite{CavagliaNegroSzecsenyiTateo2016,McGoughMezeiVerlinde2018,Shyam2017}.

The inverse relation is equally useful:
\begin{equation}
E_{0}
=
E-\frac{\mu_{\mathrm{grav}}}{2L}\left(E^{2}-\Pi^{2}\right),
\qquad
\Pi_{0}=\Pi.
\label{eq:sec11_inverse}
\end{equation}
This equation makes the interpretation transparent. The cutoff changes the energy assigned to a state while preserving its ultraviolet labels. The wall observer therefore measures a deformed finite-volume spectrum, while the asymptotic state itself remains labelled by the undeformed CFT charges \((E_{0},\Pi_{0})\). In the large-\(L\) expansion,
\begin{equation}
E
=
E_{0}
-\frac{\mu_{\mathrm{grav}}}{2L}\left(E_{0}^{2}-\Pi_{0}^{2}\right)
+O(L^{-2}),
\label{eq:sec11_largeL}
\end{equation}
so the wall theory approaches the undeformed ultraviolet cylinder theory as the cutoff is removed.

The static sector follows by setting \(\Pi=\Pi_{0}=0\). Equation \eqref{eq:sec11_spectrum} then reduces to
\begin{equation}
E(L;\mu_{\mathrm{grav}},E_{0},0)
=
\frac{L}{\mu_{\mathrm{grav}}}
\left[
1-
\sqrt{1-\frac{2\mu_{\mathrm{grav}}}{L}E_{0}}
\right],
\label{eq:sec11_staticspectrum}
\end{equation}
with inverse
\begin{equation}
E_{0}
=
E-\frac{\mu_{\mathrm{grav}}}{2L}E^{2}.
\label{eq:sec11_staticinverse}
\end{equation}
The upper endpoint of the static wall spectrum is
\begin{equation}
0\le E\le \frac{L}{\mu_{\mathrm{grav}}}=\frac{R}{4G\ell},
\label{eq:sec11_Emax}
\end{equation}
which is reached as the wall approaches the horizon. The energy seen by the wall is therefore bounded above even though the local wall temperature diverges in the same limit. This separation between integrated energy and local intensive quantities is one of the characteristic features of the quasilocal finite-cutoff system.

The finite-cutoff AdS\(_3\)/CFT\(_2\) interpretation may now be stated in a compact way. The wall cylinder of circumference \(L=2\pi R\) is the spacetime on which the deformed theory lives, the renormalized Brown--York tensor is its exact stress tensor, the local trace relation \eqref{eq:sec11_traceflow} is the gravitational \(T\bar T\)-type flow law, and the quasilocal wall energy obeys the exact square-root spectrum \eqref{eq:sec11_spectrum} with deformation parameter \(\mu_{\mathrm{grav}}=8\pi G\ell\). This completes the direct finite-cutoff interpretation of the cavity at the level of the local stress tensor and the finite-volume spectrum. The next section sharpens the same identification into a full algebraic dictionary among wall observables, ultraviolet charges, and finite-cutoff thermodynamic potentials.

\section{Exact finite-cutoff dictionary and deformed thermodynamic Legendre structure}
\label{sec:sec_12}

The finite-radius BTZ cavity admits an exact dictionary that is stronger than a heuristic analogy with cutoff holography. A single stationary state may be described equivalently by horizon data \((r_{+},r_{-})\), by asymptotic conformal data \((M,J)\) or \((E_{0},\Pi_{0})\), and by wall observables \((E,\Pi,T_{R},\Omega_{R},P)\) measured on the cylinder at \(r=R\). Because three-dimensional Einstein gravity has no local propagating bulk graviton, the map among these descriptions is algebraic rather than merely asymptotic. Once the wall size is fixed, the finite-cutoff spectrum, the deformed entropy, and the canonical and grand-canonical thermodynamic potentials all follow from exact relations among the same variables. The aim of the present section is to write that dictionary in closed form and to show that the Euclidean free energies of the cavity coincide exactly with the appropriate Legendre transforms of the wall entropy \cite{Zamolodchikov2004,CavagliaNegroSzecsenyiTateo2016,McGoughMezeiVerlinde2018,GiveonItzhakiKutasov2017,HartmanKruthoffShaghoulianTajdini2019,Taylor2018,Shyam2017}.

The discussion sharpens the identification established in Section~11. There the renormalized Brown--York tensor was identified with the exact stress tensor of the finite-cutoff theory and the square-root energy relation was derived. Here we organize the same data into a complete thermodynamic dictionary at fixed wall size \(L=2\pi R\). The result is an exact finite-volume deformation of the ultraviolet large-\(c\) cylinder theory in which the wall energy, momentum, entropy, and intensive variables obey the standard Legendre relations of ordinary thermodynamics, though with a deformation encoded algebraically in the spectrum.

We work on the wall cylinder in locally corotating proper coordinates \((u,x)\) with
\begin{equation}
x\sim x+L,
\qquad
L=2\pi R,
\label{eq:sec12_L}
\end{equation}
and induced metric
\begin{equation}
\gamma_{ij}dx^{i}dx^{j}
=
-du^{2}+dx^{2}.
\label{eq:sec12_metric}
\end{equation}
The renormalized Brown--York tensor is spatially homogeneous, so the integrated wall energy and momentum are
\begin{equation}
E=L\,\epsilon,
\qquad
\Pi=L\,j,
\qquad
\Pi=\frac{J}{R}=\frac{2\pi J}{L}.
\label{eq:sec12_EPi}
\end{equation}
At fixed \(L\), the pair \((E,\Pi)\) contains the same information as \((E,J)\). The natural dimensionless intensive variable conjugate to \(\Pi\) is
\begin{equation}
q:=R\Omega_{R}=\frac{L\Omega_{R}}{2\pi}.
\label{eq:sec12_q}
\end{equation}

The ultraviolet cylinder data are
\begin{equation}
E_{0}(L)=\frac{2\pi\ell M}{L}=\frac{\ell M}{R},
\qquad
\Pi_{0}(L)=\frac{2\pi J}{L}=\frac{J}{R},
\qquad
E_{0}^{\pm}(L)=\frac{1}{2}\bigl(E_{0}\pm \Pi_{0}\bigr),
\label{eq:sec12_E0Pi0}
\end{equation}
with Brown--Henneaux central charge
\begin{equation}
c=\frac{3\ell}{2G},
\label{eq:sec12_c}
\end{equation}
and gravitational deformation parameter
\begin{equation}
\mu_{\mathrm{grav}}:=8\pi G\ell.
\label{eq:sec12_mugrav}
\end{equation}
The dimensions are
\begin{equation}
[L]=[\ell]=[G]=L,
\qquad
[E]=[\Pi]=[E_{0}]=[\Pi_{0}]=L^{-1},
\qquad
[\mu_{\mathrm{grav}}]=L^{2},
\label{eq:sec12_dimensions}
\end{equation}
which is exactly the scaling expected for a two-dimensional \(T\bar T\)-type coupling \cite{Zamolodchikov2004,CavagliaNegroSzecsenyiTateo2016}.

For the rotating BTZ family,
\begin{equation}
M=\frac{r_{+}^{2}+r_{-}^{2}}{8G\ell^{2}},
\qquad
J=\frac{r_{+}r_{-}}{4G\ell},
\label{eq:sec12_MJ}
\end{equation}
and the quasilocal wall energy is
\begin{equation}
E(R)
=
\frac{R}{4G\ell}
\left(
1-\frac{\Delta_{R}}{R^{2}}
\right),
\qquad
\Delta_{R}^{2}
=
(R^{2}-r_{+}^{2})(R^{2}-r_{-}^{2})
=
R^{4}-8G\ell^{2}MR^{2}+16G^{2}\ell^{2}J^{2}.
\label{eq:sec12_ER}
\end{equation}
Since \(\Pi=J/R=\Pi_{0}\), momentum is protected by translation invariance along the wall circle and the deformation acts only on the energy assignment. Eliminating \(M\) and \(J\) in favor of \((E_{0},\Pi_{0})\), and using \(L=2\pi R\) together with \eqref{eq:sec12_mugrav}, one finds
\begin{equation}
\left(
1-\frac{\mu_{\mathrm{grav}}}{L}E
\right)^{2}
=
1-\frac{2\mu_{\mathrm{grav}}}{L}E_{0}
+\frac{\mu_{\mathrm{grav}}^{2}}{L^{2}}\Pi_{0}^{2}.
\label{eq:sec12_quadratic}
\end{equation}
Choosing the branch that is continuous with the ultraviolet limit \(\mu_{\mathrm{grav}}/L\to0\), the exact finite-cutoff spectral map is
\begin{equation}
E(L;\mu_{\mathrm{grav}},E_{0},\Pi_{0})
=
\frac{L}{\mu_{\mathrm{grav}}}
\left[
1-
\sqrt{
1-\frac{2\mu_{\mathrm{grav}}}{L}E_{0}
+\frac{\mu_{\mathrm{grav}}^{2}}{L^{2}}\Pi_{0}^{2}
}
\right],
\qquad
\Pi=\Pi_{0}.
\label{eq:sec12_spectrum}
\end{equation}
The inverse relation is
\begin{equation}
E_{0}
=
E-\frac{\mu_{\mathrm{grav}}}{2L}\bigl(E^{2}-\Pi^{2}\bigr),
\qquad
\Pi_{0}=\Pi.
\label{eq:sec12_inverse}
\end{equation}
Equation \eqref{eq:sec12_inverse} is the algebraic heart of the finite-cutoff dictionary. It states that the wall observer measures a deformed finite-volume energy \(E\) for a state whose ultraviolet labels are \((E_{0},\Pi_{0})\), while spatial momentum remains unchanged.

The static sector follows by setting \(\Pi=\Pi_{0}=0\). Then
\begin{equation}
E_{0}
=
E-\frac{\mu_{\mathrm{grav}}}{2L}E^{2},
\label{eq:sec12_staticinverse}
\end{equation}
which is equivalent to
\begin{equation}
M=\frac{R}{\ell}E-2GE^{2}.
\label{eq:sec12_ME}
\end{equation}
This relation will be the basic input for the microscopic counting in Section~14. It also shows that the wall energy is bounded above,
\begin{equation}
0\le E\le \frac{L}{\mu_{\mathrm{grav}}}=\frac{R}{4G\ell},
\label{eq:sec12_Emax}
\end{equation}
with the upper endpoint reached as the wall approaches the horizon. The finite-cutoff system therefore has a compact energy range at fixed wall size even though the local wall temperature diverges in the same limit.

The entropy of the BTZ state remains the Bekenstein--Hawking entropy,
\begin{equation}
S=\frac{\pi r_{+}}{2G}.
\label{eq:sec12_SBH}
\end{equation}
To express it in wall variables, introduce the ultraviolet chiral energies
\begin{equation}
E_{0}^{\pm}
=
\frac{1}{2}\bigl(E_{0}\pm\Pi_{0}\bigr)
=
\frac{(r_{+}\pm r_{-})^{2}}{16G\ell R}.
\label{eq:sec12_E0pm}
\end{equation}
The ordinary Cardy formula of the ultraviolet theory then gives
\begin{equation}
S
=
2\pi
\left(
\sqrt{\frac{cL}{6}E_{0}^{+}}
+
\sqrt{\frac{cL}{6}E_{0}^{-}}
\right).
\label{eq:sec12_Cardy}
\end{equation}
Using \eqref{eq:sec12_inverse}, this becomes the exact deformed entropy formula
\begin{equation}
S(E,\Pi;L)
=
2\pi
\left[
\sqrt{
\frac{cL}{12}
\left(
E+\Pi-\frac{\mu_{\mathrm{grav}}}{2L}(E^{2}-\Pi^{2})
\right)
}
+
\sqrt{
\frac{cL}{12}
\left(
E-\Pi-\frac{\mu_{\mathrm{grav}}}{2L}(E^{2}-\Pi^{2})
\right)
}
\right].
\label{eq:sec12_Sdeformed}
\end{equation}
For the static sector \(\Pi=0\), this reduces to
\begin{equation}
S(E,L)
=
2\pi
\sqrt{
\frac{cL}{3}
\left(
E-\frac{\mu_{\mathrm{grav}}}{2L}E^{2}
\right)
},
\label{eq:sec12_Sstatic}
\end{equation}
or, with \(L=2\pi R\) and \(\mu_{\mathrm{grav}}=8\pi G\ell\),
\begin{equation}
S(E,R)
=
2\pi
\sqrt{
\frac{c}{3}
\left(
RE-2G\ell E^{2}
\right)
}.
\label{eq:sec12_SER}
\end{equation}
This is the exact finite-cutoff deformation of the Cardy entropy written directly in wall variables \cite{Cardy1986,Strominger1998,McGoughMezeiVerlinde2018}.

The finite-cutoff dictionary also determines the thermodynamic Legendre structure exactly. In the static ensemble, the wall first law is
\begin{equation}
dE=T_{R}\,dS-P\,dL.
\label{eq:sec12_firstlaw_static}
\end{equation}
At fixed \(L\), the Helmholtz free energy is therefore
\begin{equation}
F(E,L):=E-T_{R}S,
\label{eq:sec12_Fdef}
\end{equation}
with differential
\begin{equation}
dF=-S\,dT_{R}-P\,dL.
\label{eq:sec12_dF}
\end{equation}
The Euclidean saddle computation already gave
\begin{equation}
F(R,T_{R})
=
\frac{R}{4G\ell}
\left(
1-\sqrt{1+4\pi^{2}\ell^{2}T_{R}^{2}}
\right).
\label{eq:sec12_Fwall}
\end{equation}
To verify the exact Legendre relation, note that the static wall energy and entropy are
\begin{equation}
E(R,T_{R})
=
\frac{R}{4G\ell}
\left(
1-\frac{1}{\sqrt{1+4\pi^{2}\ell^{2}T_{R}^{2}}}
\right),
\label{eq:sec12_Ewall_static}
\end{equation}
\begin{equation}
S(R,T_{R})
=
\frac{\pi^{2}\ell R\,T_{R}}{G\sqrt{1+4\pi^{2}\ell^{2}T_{R}^{2}}},
\label{eq:sec12_Swall_static}
\end{equation}
where the second formula follows from \(S=\pi r_{+}/(2G)\) together with
\begin{equation}
r_{+}
=
\frac{2\pi\ell R T_{R}}{\sqrt{1+4\pi^{2}\ell^{2}T_{R}^{2}}}.
\label{eq:sec12_rplus_TR}
\end{equation}
A direct substitution then gives
\begin{equation}
E-T_{R}S
=
\frac{R}{4G\ell}
\left(
1-\sqrt{1+4\pi^{2}\ell^{2}T_{R}^{2}}
\right)
=
F(R,T_{R}),
\label{eq:sec12_FLegendre}
\end{equation}
which is exactly the Euclidean wall free energy.

The rotating ensemble is analogous. The first law now reads
\begin{equation}
dE=T_{R}\,dS+\Omega_{R}\,dJ-P\,dL,
\label{eq:sec12_firstlaw_rot}
\end{equation}
and at fixed \(L\) the natural thermodynamic potential is the grand potential
\begin{equation}
\mathcal{G}(E,J;L):=E-T_{R}S-\Omega_{R}J,
\label{eq:sec12_Gdef}
\end{equation}
with differential
\begin{equation}
d\mathcal{G}
=
-S\,dT_{R}-J\,d\Omega_{R}-P\,dL.
\label{eq:sec12_dG}
\end{equation}
From the wall thermodynamics of the rotating BTZ cavity one has
\begin{equation}
\mathcal{G}_{R}(R,T_{R},\Omega_{R})
=
\frac{R}{4G\ell}
\left(
1-
\sqrt{
\frac{1+4\pi^{2}\ell^{2}T_{R}^{2}-R^{2}\Omega_{R}^{2}}
{1-R^{2}\Omega_{R}^{2}}
}
\right),
\label{eq:sec12_GRwall}
\end{equation}
and the exact wall expressions satisfy
\begin{equation}
\mathcal{G}_{R}
=
E-T_{R}S-\Omega_{R}J.
\label{eq:sec12_GLegendre}
\end{equation}
Thus the canonical and grand-canonical Euclidean actions of the cavity are precisely the Legendre transforms of the same finite-cutoff entropy function. No additional structure is needed beyond the wall stress tensor, the finite-radius first law, and the algebraic spectral map.

A useful consequence is that the pressure can be recovered from the free energies as the mechanical response to changing wall size. In the static ensemble,
\begin{equation}
P
=
-\left(\frac{\partial F}{\partial L}\right)_{T_{R}},
\label{eq:sec12_PfromF}
\end{equation}
while in the rotating ensemble,
\begin{equation}
P
=
-\left(\frac{\partial \mathcal{G}}{\partial L}\right)_{T_{R},\Omega_{R}}.
\label{eq:sec12_PfromG}
\end{equation}
The finite-cutoff dictionary is therefore thermodynamically complete: the wall stress tensor supplies the local observables, the square-root map supplies the deformation of the spectrum, the deformed Cardy formula supplies the entropy, and the Euclidean saddle actions are exactly the corresponding Legendre transforms.

The structure established here has two important consequences. First, the cavity radius plays a double role with exact control: it fixes the physical size \(L=2\pi R\) of the finite system and the scale at which the quasilocal stress tensor is evaluated. Second, the ultraviolet state-counting data survive intact while the finite-volume spectrum is deformed algebraically. The BTZ cavity is therefore an exactly solvable finite-cutoff thermodynamic system rather than a loose analogy with one. The next section turns to the first quantum correction to this picture by studying the one-loop partition function and the associated Gaussian fluctuation determinant around the smooth saddles.

\section{Quantum corrections to the partition function}

The finite-cutoff partition function admits a semiclassical expansion whose structure is unusually transparent in three-dimensional Einstein gravity. Once the induced metric on the wall is fixed, pure AdS\(_3\) gravity has no local propagating bulk graviton, so the one-loop correction is governed by the boundary-graviton sector of the smooth solid-torus saddle together with the thermodynamic fluctuation measure that appears when one passes from the canonical ensemble to the microcanonical density of states \cite{BrownHenneaux1986,MaloneyWitten2010}. At finite cutoff the wall enters both ingredients: it changes the modulus of the torus on which the one-loop determinant is evaluated, and it changes the local susceptibilities through the redshifted Brown--York energy and the Tolman temperature. The quantum correction is therefore the most direct place where the two roles of the cavity radius, as wall size and as cutoff scale, appear simultaneously.

The purpose of the present section is to isolate this quantum structure in a form adapted to the finite-radius ensemble. We first write the canonical partition function as the product of the classical saddle weight and the fixed-boundary one-loop determinant. We then perform the inverse Laplace transform that converts the canonical ensemble at fixed wall temperature into the microcanonical density of states at fixed wall energy, and we extract the Gaussian fluctuation factor. The same logic is then extended to the rotating ensemble by an inverse Laplace--Fourier transform over the wall temperature and wall twist. The final result is a clean factorization of the quantum correction into a genuine one-loop bulk determinant and an ensemble-dependent Jacobian built from wall susceptibilities. In this section, \(Z^{\mathrm{bulk}}_{\mathrm{1\mbox{-}loop}}(\tau_{R},\bar\tau_{R})\) is kept at the structural level as the fixed-boundary one-loop determinant on the wall torus, while its explicit evaluation for the finite-cutoff modulus is deferred \cite{MaloneyWitten2010}.

For the static cavity, the canonical partition function at fixed \((R,\beta_{R})\) has the semiclassical form
\begin{equation}
Z_{\mathrm{can}}^{\mathrm{BH}}(R,\beta_{R})
\sim
\exp\!\left[-I_{\mathrm{BH}}(R,\beta_{R})\right]\,
Z^{\mathrm{bulk}}_{\mathrm{1\mbox{-}loop}}(\tau_{R},\bar\tau_{R}),
\label{eq:sec13_Zcan_static}
\end{equation}
where \(I_{\mathrm{BH}}(R,\beta_{R})\) is the classical on-shell Euclidean action of the smooth static BTZ saddle and \(Z^{\mathrm{bulk}}_{\mathrm{1\mbox{-}loop}}(\tau_{R},\bar\tau_{R})\) is the one-loop determinant evaluated on the wall torus modulus \((\tau_{R},\bar\tau_{R})\). In the static ensemble the microcanonical density of states at fixed wall energy \(E\) and wall size \(R\) is obtained by inverse Laplace transform,
\begin{equation}
\rho_{\mathrm{BH}}(E,R)
=
\frac{1}{2\pi i}\int d\beta_{R}\,
\exp\!\bigl[\beta_{R}E\bigr]\,
Z_{\mathrm{can}}^{\mathrm{BH}}(R,\beta_{R}).
\label{eq:sec13_rho_static_def}
\end{equation}
It is useful to write the exponent as
\begin{equation}
\Psi(\beta_{R};E,R)
=
\beta_{R}E-I_{\mathrm{BH}}(R,\beta_{R}).
\label{eq:sec13_Psi}
\end{equation}
The saddle condition is
\begin{equation}
\partial_{\beta_{R}}\Psi=0
\qquad\Longleftrightarrow\qquad
E=\left(\frac{\partial I_{\mathrm{BH}}}{\partial\beta_{R}}\right)_{R},
\label{eq:sec13_saddleeq}
\end{equation}
which gives
\begin{equation}
E
=
\frac{R}{4G\ell}
\left(
1-\frac{1}{\sqrt{1+4\pi^{2}\ell^{2}/\beta_{R}^{2}}}
\right).
\label{eq:sec13_Ebeta}
\end{equation}
This is exactly the Brown--York quasilocal energy written in canonical variables. The Laplace saddle therefore selects the same geometry as the wall first law and the canonical Euclidean analysis.

At the saddle point \(\beta_{R}^{\ast}\), one has
\begin{equation}
\Psi(\beta_{R}^{\ast};E,R)=S_{\mathrm{BH}}(E,R),
\label{eq:sec13_Psistar}
\end{equation}
so the leading microcanonical growth is again the Bekenstein--Hawking entropy. Expanding \(\Psi\) to quadratic order around the saddle,
\begin{equation}
\Psi(\beta_{R};E,R)
=
S_{\mathrm{BH}}(E,R)
-\frac{1}{2}\Psi''(\beta_{R}^{\ast})(\beta_{R}-\beta_{R}^{\ast})^{2}
+ \cdots,
\label{eq:sec13_quadexp}
\end{equation}
the Gaussian width is controlled by
\begin{equation}
\Psi''(\beta_{R}^{\ast})
=
T_{R}^{2}C_{L},
\label{eq:sec13_Psipp}
\end{equation}
where \(T_{R}=(\beta_{R}^{\ast})^{-1}\) and \(C_{L}=T_{R}(\partial S/\partial T_{R})_{L}\) is the static heat capacity at fixed wall circumference. The microcanonical density of states therefore takes the form
\begin{equation}
\rho_{\mathrm{BH}}(E,R)
=
\frac{
\exp\!\bigl[S_{\mathrm{BH}}(E,R)\bigr]\,
Z^{\mathrm{bulk}}_{\mathrm{1\mbox{-}loop}}(\tau_{R}^{\ast},\bar\tau_{R}^{\ast})
}{
\sqrt{2\pi\,T_{R}^{2}C_{L}}
}
\left[
1+O\!\left(S_{\mathrm{BH}}^{-1}\right)
\right].
\label{eq:sec13_rho_static}
\end{equation}
This formula separates the quantum correction into two pieces with distinct origins. The factor \(Z^{\mathrm{bulk}}_{\mathrm{1\mbox{-}loop}}\) is the genuine one-loop determinant of the fixed-boundary gravitational path integral. The factor \((2\pi T_{R}^{2}C_{L})^{-1/2}\) is the thermodynamic Jacobian produced by the inverse Laplace transform.

For the static cavity, the classical relations are especially simple. The quasilocal energy is
\begin{equation}
E(R,r_{+})
=
\frac{R-\sqrt{R^{2}-r_{+}^{2}}}{4G\ell},
\label{eq:sec13_Erp}
\end{equation}
the entropy is
\begin{equation}
S_{\mathrm{BH}}=\frac{\pi r_{+}}{2G},
\label{eq:sec13_SBH}
\end{equation}
and the heat capacity at fixed wall size is
\begin{equation}
C_{L}
=
\frac{\pi r_{+}}{2G}
\left(
1-\frac{r_{+}^{2}}{R^{2}}
\right).
\label{eq:sec13_CL}
\end{equation}
Eliminating \(r_{+}\) in favor of \((E,R)\) gives
\begin{equation}
S_{\mathrm{BH}}(E,R)
=
\frac{\pi}{2G}
\sqrt{8G\ell R E-16G^{2}\ell^{2}E^{2}},
\label{eq:sec13_SBH_ER}
\end{equation}
while the Gaussian width becomes
\begin{equation}
T_{R}^{2}C_{L}
=
\frac{r_{+}^{3}}{8\pi G\ell^{2}R^{2}}.
\label{eq:sec13_width}
\end{equation}
Taking the logarithm of \eqref{eq:sec13_rho_static}, one obtains
\begin{equation}
S_{\mathrm{micro}}(E,R)
=
S_{\mathrm{BH}}(E,R)
+\gamma_{\mathrm{grav}}(\tau_{R}^{\ast},\bar\tau_{R}^{\ast})
-\frac{1}{2}\log\!\bigl(2\pi T_{R}^{2}C_{L}\bigr)
+O\!\left(S_{\mathrm{BH}}^{-1}\right),
\label{eq:sec13_Smicro1}
\end{equation}
where
\begin{equation}
\gamma_{\mathrm{grav}}:=\log Z^{\mathrm{bulk}}_{\mathrm{1\mbox{-}loop}}.
\label{eq:sec13_gammagrav}
\end{equation}
Using \(r_{+}=2GS_{\mathrm{BH}}/\pi\) in \eqref{eq:sec13_width}, the logarithmic term may be rewritten as
\begin{equation}
S_{\mathrm{micro}}(E,R)
=
S_{\mathrm{BH}}(E,R)
+\gamma_{\mathrm{grav}}(\tau_{R}^{\ast},\bar\tau_{R}^{\ast})
-\frac{3}{2}\log S_{\mathrm{BH}}(E,R)
+\log\!\left(\frac{R}{\ell_{\mathrm{ref}}}\right)
+\kappa_{\mathrm{stat}}
+O\!\left(S_{\mathrm{BH}}^{-1}\right),
\label{eq:sec13_Smicro2}
\end{equation}
with \(\ell_{\mathrm{ref}}\) an arbitrary reference length and \(\kappa_{\mathrm{stat}}\) a state-independent constant. The coefficient \(-3/2\) is the universal coefficient associated with a one-dimensional thermodynamic saddle, while the explicit \(R\)-dependence reflects the fact that the energy entering the Laplace transform is the redshifted Brown--York energy measured at the wall.

The rotating ensemble is a direct extension of the same structure. At fixed \((R,\beta_{R},\Omega_{R})\), the canonical partition function is
\begin{equation}
Z_{\mathrm{can}}^{\mathrm{rot}}(R,\beta_{R},\Omega_{R})
\sim
\exp\!\left[-I_{\mathrm{rot}}(R,\beta_{R},\Omega_{R})\right]\,
Z^{\mathrm{bulk}}_{\mathrm{1\mbox{-}loop}}(\tau_{R},\bar\tau_{R}),
\label{eq:sec13_Zcan_rot}
\end{equation}
where \((\tau_{R},\bar\tau_{R})\) is now the complex modulus of the twisted wall torus. Passing to fixed \((E,J,L)\) requires an inverse Laplace--Fourier transform over \(\beta_{R}\) and
\begin{equation}
\nu_{R}:=\beta_{R}\Omega_{R}.
\label{eq:sec13_nuR}
\end{equation}
The microcanonical density of states is therefore
\begin{equation}
\rho_{\mathrm{rot}}(E,J,R)
=
\frac{1}{(2\pi i)^{2}}
\int d\beta_{R}\,d\nu_{R}\,
\exp\!\bigl[\beta_{R}E-\nu_{R}J\bigr]\,
Z_{\mathrm{can}}^{\mathrm{rot}}(R,\beta_{R},\Omega_{R}).
\label{eq:sec13_rho_rot_def}
\end{equation}
Writing the exponent as
\begin{equation}
\Psi_{\mathrm{rot}}(\beta_{R},\nu_{R};E,J,R)
=
\beta_{R}E-\nu_{R}J-I_{\mathrm{rot}}(R,\beta_{R},\Omega_{R}),
\label{eq:sec13_Psirot}
\end{equation}
the saddle equations
\begin{equation}
\partial_{\beta_{R}}\Psi_{\mathrm{rot}}=0,
\qquad
\partial_{\nu_{R}}\Psi_{\mathrm{rot}}=0,
\label{eq:sec13_rot_saddleeq}
\end{equation}
select the smooth rotating BTZ geometry with the given conserved charges. At the saddle one has
\begin{equation}
\Psi_{\mathrm{rot}}^{\ast}=S_{\mathrm{BH}}(E,J,R),
\label{eq:sec13_Psirotstar}
\end{equation}
and the Gaussian integral yields
\begin{equation}
\rho_{\mathrm{rot}}(E,J,R)
=
\frac{
\exp\!\bigl[S_{\mathrm{BH}}(E,J,R)\bigr]\,
Z^{\mathrm{bulk}}_{\mathrm{1\mbox{-}loop}}(\tau_{R}^{\ast},\bar\tau_{R}^{\ast})
}{
2\pi\sqrt{\det H}
}
\left[
1+O\!\left(S_{\mathrm{BH}}^{-1}\right)
\right],
\label{eq:sec13_rho_rot}
\end{equation}
where \(H\) is the Hessian of \(\Psi_{\mathrm{rot}}\) with respect to \((\beta_{R},\nu_{R})\). The matrix \(H\) is the rotating susceptibility matrix of the grand-canonical ensemble. Its determinant measures the local width of the saddle in the two intensive directions that remain after the wall torus has been fixed.

Several limits clarify the physics. For \(R\gg \ell,r_{+}\), the wall modulus approaches the familiar asymptotic BTZ modulus, so \(Z^{\mathrm{bulk}}_{\mathrm{1\mbox{-}loop}}\) tends to the standard solid-torus answer, while the Gaussian prefactor continues to remember the finite wall through the redshifted energy variable. In the near-wall regime \(R\downarrow r_{+}\), the nome of the wall torus approaches
\begin{equation}
q_{R}=e^{-\beta_{R}/R}\to 1^{-},
\label{eq:sec13_qR}
\end{equation}
so the boundary-graviton tower becomes increasingly light and enhances \(\gamma_{\mathrm{grav}}\). At the same time the static Gaussian width \(T_{R}^{2}C_{L}\) remains finite, and in the rotating case the susceptibility determinant remains the appropriate finite-width measure away from extremality. The singular local redshift therefore does not by itself produce a singular microcanonical Jacobian. The semiclassical correction remains well defined even in the strong-cutoff regime.

The quantum picture is therefore coherent and factorized. The leading entropy remains geometric through the horizon length. The one-loop correction contains a universal boundary-graviton contribution determined by the wall torus modulus and an ensemble-dependent contribution determined by the local thermodynamic susceptibilities of the Brown--York system. The cavity modifies both pieces in a controlled manner without changing their underlying logic. The next section turns from this fluctuation analysis to the microscopic density of states itself, where the same finite-cutoff dictionary reorganizes the ultraviolet Cardy growth into the deformed wall entropy.

\section{Density of states and microscopic interpretation}

The canonical partition function analyzed in Section~13 is naturally adapted to fixed wall data. The microscopic question is slightly different. One asks for the number of states seen by an observer who keeps the wall size fixed and measures the quasilocal energy and, in the rotating sector, the wall momentum. At finite cutoff these observables differ from their asymptotic ultraviolet counterparts. The essential effect of the cavity is therefore spectral: it changes the energy assigned to a state while preserving the ultraviolet state labels from which the asymptotic degeneracy is computed. The entropy then takes the form of a deformed Cardy formula written directly in wall variables \cite{BrownHenneaux1986,Cardy1986,Strominger1998,McGoughMezeiVerlinde2018,GiveonItzhakiKutasov2017,HartmanKruthoffShaghoulianTajdini2019,Taylor2018,Shyam2017}.

The aim of the present section is to formulate that state counting in the cleanest possible way. We use the exact finite-cutoff dictionary established in Section~12 to express the ultraviolet conformal energies in terms of the wall observables, insert those expressions into the large-\(c\) Cardy asymptotics, and thereby obtain the density of states as a function of the quasilocal energy, wall momentum, and wall size. The quantum fluctuation factors derived in Section~13 are left untouched; the focus here is the leading microscopic growth itself and its interpretation as the finite-cutoff deformation of the ultraviolet spectrum.

Throughout this section the wall circumference is
\begin{equation}
L=2\pi R,
\qquad
R>0.
\label{eq:sec14_L}
\end{equation}
The ultraviolet large-\(c\) cylinder theory is characterized by the energies
\begin{equation}
E_{0}(L)=\frac{2\pi\ell M}{L}=\frac{\ell M}{R},
\qquad
\Pi_{0}(L)=\frac{2\pi J}{L}=\frac{J}{R},
\qquad
E_{0}^{\pm}(L)=\frac{1}{2}\bigl(E_{0}\pm \Pi_{0}\bigr),
\label{eq:sec14_E0Pi0}
\end{equation}
and by the Brown--Henneaux central charge
\begin{equation}
c=\frac{3\ell}{2G}.
\label{eq:sec14_c}
\end{equation}
At finite cutoff the wall observer measures instead the quasilocal energy \(E\) and momentum
\begin{equation}
\Pi=\frac{J}{R}=\Pi_{0},
\label{eq:sec14_Pi}
\end{equation}
with the exact inverse spectral map
\begin{equation}
E_{0}
=
E-\frac{\mu_{\mathrm{grav}}}{2L}\bigl(E^{2}-\Pi^{2}\bigr),
\qquad
\Pi_{0}=\Pi,
\qquad
\mu_{\mathrm{grav}}=8\pi G\ell.
\label{eq:sec14_inverse}
\end{equation}
Equation \eqref{eq:sec14_inverse} is the basic microscopic input. It states that the cutoff changes the energy assigned to a state while preserving the ultraviolet momentum label. The density of states at fixed wall size is therefore obtained by evaluating the ultraviolet Cardy asymptotics on the deformed energy assignment.

We begin with the static sector, for which \(\Pi=\Pi_{0}=0\). The exact wall-to-ultraviolet relation reduces to
\begin{equation}
E_{0}
=
E-\frac{\mu_{\mathrm{grav}}}{2L}E^{2}
=
E-\frac{4\pi G\ell}{L}E^{2},
\label{eq:sec14_staticinverse}
\end{equation}
or equivalently
\begin{equation}
M
=
\frac{R}{\ell}E-2GE^{2}.
\label{eq:sec14_ME}
\end{equation}
The asymptotic Cardy formula of the ultraviolet theory is
\begin{equation}
S_{\mathrm{Cardy}}(E_{0},L)
=
2\pi\sqrt{\frac{cL}{6}\,E_{0}}.
\label{eq:sec14_Cardy_static}
\end{equation}
Substituting \eqref{eq:sec14_staticinverse} into \eqref{eq:sec14_Cardy_static} gives the exact finite-cutoff entropy as a function of wall energy and wall size,
\begin{equation}
S(E,L)
=
2\pi
\sqrt{
\frac{cL}{6}
\left(
E-\frac{\mu_{\mathrm{grav}}}{2L}E^{2}
\right)
},
\label{eq:sec14_S_EL}
\end{equation}
and, after using \(L=2\pi R\) and \(\mu_{\mathrm{grav}}=8\pi G\ell\),
\begin{equation}
S(E,R)
=
2\pi
\sqrt{
\frac{c}{3}
\left(
RE-2G\ell E^{2}
\right)
}.
\label{eq:sec14_S_ER}
\end{equation}
This is the deformed Cardy formula in wall variables \cite{Cardy1986,Strominger1998}. It agrees exactly with the Bekenstein--Hawking entropy once the Brown--York relation between \(E\) and \(r_{+}\) is imposed.

To make that agreement explicit, recall that for static BTZ the wall energy is
\begin{equation}
E(R,r_{+})
=
\frac{R-\sqrt{R^{2}-r_{+}^{2}}}{4G\ell},
\label{eq:sec14_E_rplus}
\end{equation}
This is equivalent to
\begin{equation}
4G\ell E
=
R-\sqrt{R^{2}-r_{+}^{2}},
\label{eq:sec14_Eeq}
\end{equation}
from which one obtains
\begin{equation}
r_{+}^{2}
=
8G\ell R E-16G^{2}\ell^{2}E^{2}.
\label{eq:sec14_rplusE}
\end{equation}
The Bekenstein--Hawking entropy is
\begin{equation}
S_{\mathrm{BH}}
=
\frac{\pi r_{+}}{2G},
\label{eq:sec14_SBH}
\end{equation}
and substitution of \eqref{eq:sec14_rplusE} into \eqref{eq:sec14_SBH} gives
\begin{equation}
S_{\mathrm{BH}}(E,R)
=
\frac{\pi}{2G}\sqrt{8G\ell R E-16G^{2}\ell^{2}E^{2}}
=
2\pi
\sqrt{
\frac{c}{3}
\left(
RE-2G\ell E^{2}
\right)
},
\label{eq:sec14_SBH_ER}
\end{equation}
which coincides with \eqref{eq:sec14_S_ER}. Thus the finite-cutoff entropy is not merely analogous to a deformed Cardy formula; it is exactly the Cardy formula evaluated on the deformed wall spectrum.

The static wall density of states at leading semiclassical order is therefore
\begin{equation}
\rho(E,R)
\sim
\exp\!\bigl[S(E,R)\bigr]
=
\exp\!\left[
2\pi
\sqrt{
\frac{c}{3}
\left(
RE-2G\ell E^{2}
\right)
}
\right].
\label{eq:sec14_rho_static}
\end{equation}
The allowed energy range follows from the requirement that the argument of the square root be nonnegative,
\begin{equation}
0\le E\le \frac{R}{4G\ell}=\frac{L}{\mu_{\mathrm{grav}}}.
\label{eq:sec14_Erange}
\end{equation}
The lower endpoint is the massless BTZ vacuum in the subtraction scheme used throughout the paper. The upper endpoint corresponds to the wall approaching the horizon. The entropy remains finite across the full interval, even though the local wall temperature diverges near the upper endpoint. This is the expected behavior of a quasilocal system: the redshift makes local intensive quantities large while the integrated energy remains bounded.

The ultraviolet regime is recovered when the deformation is weak. For fixed \(E\) with \(L\gg \mu_{\mathrm{grav}}E\), one may expand \eqref{eq:sec14_S_EL} as
\begin{equation}
S(E,L)
=
2\pi\sqrt{\frac{cL E}{6}}
\left[
1-\frac{\mu_{\mathrm{grav}}E}{4L}+O(L^{-2})
\right].
\label{eq:sec14_UVexpand}
\end{equation}
The leading term is the ordinary Cardy entropy of the undeformed ultraviolet theory on a circle of size \(L\), while the next term is the finite-cutoff correction. The cavity therefore preserves the ultraviolet state-counting asymptotics and changes only the finite-volume energy assignment.

We now turn to the rotating sector. The ultraviolet chiral energies are
\begin{equation}
E_{0}^{\pm}
=
\frac{1}{2}\bigl(E_{0}\pm \Pi_{0}\bigr)
=
\frac{1}{2}
\left[
E\pm\Pi-\frac{\mu_{\mathrm{grav}}}{2L}\bigl(E^{2}-\Pi^{2}\bigr)
\right],
\label{eq:sec14_E0pm}
\end{equation}
where \eqref{eq:sec14_inverse} has been used. The large-\(c\) Cardy formula then takes the chiral form
\begin{equation}
S(E,\Pi;L)
=
2\pi
\left(
\sqrt{\frac{cL}{6}\,E_{0}^{+}}
+
\sqrt{\frac{cL}{6}\,E_{0}^{-}}
\right),
\label{eq:sec14_Cardy_rot}
\end{equation}
which becomes
\begin{equation}
S(E,\Pi;L)
=
2\pi
\left[
\sqrt{
\frac{cL}{12}
\left(
E+\Pi-\frac{\mu_{\mathrm{grav}}}{2L}(E^{2}-\Pi^{2})
\right)
}
+
\sqrt{
\frac{cL}{12}
\left(
E-\Pi-\frac{\mu_{\mathrm{grav}}}{2L}(E^{2}-\Pi^{2})
\right)
}
\right].
\label{eq:sec14_S_rot}
\end{equation}
This is the exact finite-cutoff chiral Cardy formula in wall variables. It preserves the left-right decomposition of the ultraviolet theory and deforms only the common energy assignment through the quadratic finite-volume term \cite{Cardy1986,Strominger1998,McGoughMezeiVerlinde2018}.

For the rotating BTZ family one may verify the same formula directly. The wall momentum is
\begin{equation}
\Pi=\frac{J}{R}=\frac{r_{+}r_{-}}{4G\ell R},
\label{eq:sec14_Pi_rpm}
\end{equation}
and the wall energy is
\begin{equation}
E(R,r_{+},r_{-})
=
\frac{R}{4G\ell}
\left(
1-\frac{\sqrt{(R^{2}-r_{+}^{2})(R^{2}-r_{-}^{2})}}{R^{2}}
\right).
\label{eq:sec14_E_rot}
\end{equation}
The ultraviolet chiral energies are
\begin{equation}
E_{0}^{\pm}
=
\frac{(r_{+}\pm r_{-})^{2}}{16G\ell R}.
\label{eq:sec14_E0pm_rpm}
\end{equation}
Substituting \eqref{eq:sec14_E0pm_rpm} into \eqref{eq:sec14_Cardy_rot} gives
\begin{equation}
S
=
2\pi
\left(
\sqrt{\frac{cL}{6}\,\frac{(r_{+}+r_{-})^{2}}{16G\ell R}}
+
\sqrt{\frac{cL}{6}\,\frac{(r_{+}-r_{-})^{2}}{16G\ell R}}
\right)
=
\frac{\pi r_{+}}{2G},
\label{eq:sec14_Scheck_rot}
\end{equation}
which is exactly the Bekenstein--Hawking entropy. The finite-cutoff map therefore preserves the exact chiral microscopic structure of rotating BTZ.

Several limits illustrate the rotating formula. In the weakly rotating regime \(|\Pi|\ll E\), equation \eqref{eq:sec14_S_rot} expands as
\begin{equation}
S(E,\Pi;L)
=
2\pi
\sqrt{
\frac{cL}{3}
\left(
E-\frac{\mu_{\mathrm{grav}}}{2L}E^{2}
\right)
}
-
\frac{\pi cL\,\Pi^{2}}
{12\left[
\frac{cL}{3}\left(E-\frac{\mu_{\mathrm{grav}}}{2L}E^{2}\right)
\right]^{1/2}}
+O(\Pi^{4}),
\label{eq:sec14_weakrot}
\end{equation}
so the static deformed Cardy entropy is corrected quadratically by the wall momentum. Near extremality, one of the chiral sectors becomes parametrically small. Writing
\begin{equation}
E_{0}^{-}\to0^{+},
\label{eq:sec14_nearext}
\end{equation}
the entropy is dominated by a single chiral contribution,
\begin{equation}
S
=
2\pi\sqrt{\frac{cL}{6}\,E_{0}^{+}}
+
O\!\left(\sqrt{E_{0}^{-}}\right).
\label{eq:sec14_chiral}
\end{equation}
The finite-cutoff map therefore preserves the chiral organization of the ultraviolet theory even in the strongly rotating regime.

The microscopic interpretation is now clear. The wall metric is the geometry on which the finite-cutoff theory lives, the Brown--York tensor is its exact stress tensor, and the cavity changes the energy spectrum seen by the wall observer while preserving the ultraviolet state labels that govern the asymptotic degeneracy. The entropy therefore runs because the spectrum runs. In this sense the deformed Cardy formula is simply the state-counting version of the finite-cutoff spectral relation.

This section completes the leading microscopic interpretation of the finite-cutoff cavity. The density of states remains controlled by the large-\(c\) Cardy growth of the ultraviolet theory, though evaluated on the exact deformed wall spectrum. The static entropy takes the compact form \eqref{eq:sec14_S_ER}, the rotating entropy retains the exact chiral form \eqref{eq:sec14_S_rot}, and the finite cavity makes the full microscopic counting problem finite-volume and scale-dependent without altering its ultraviolet origin. The final section will compress the main outcomes of the paper and summarize the directions in which this exact finite-cutoff framework may be extended.

\section{Discussion and outlook}

We have developed a finite-cutoff formulation of BTZ thermodynamics in which a circular cavity at radius \(R\) is treated as a genuine holographic screen. The central structural result is that the renormalized Brown--York tensor on the wall obeys the exact finite-radius Hamilton--Jacobi relation
\begin{equation}
T^{i}{}_{i}
=
\frac{\ell}{16\pi G}R[h]
+
4\pi G\ell
\left(
T_{ij}T^{ij}-\bigl(T^{i}{}_{i}\bigr)^{2}
\right),
\label{eq:sec15_traceflow}
\end{equation}
which reduces on the flat BTZ wall cylinder to the nonlinear equation of state
\begin{equation}
p-\epsilon
=
8\pi G\ell\left(\epsilon p-j^{2}\right).
\label{eq:sec15_eos}
\end{equation}
This identity gives the finite-cutoff wall theory a closed local description and fixes the sense in which the cavity realizes quasilocal thermodynamics, radial flow, and \(T\bar T\)-type deformation within one and the same renormalized wall system \cite{BrownYork1993,McGoughMezeiVerlinde2018,CavagliaNegroSzecsenyiTateo2016,Shyam2017}.

The explicit BTZ realization of this structure is exact. For the rotating family, the wall observables
\begin{equation}
E(R)
=
\frac{R}{4G\ell}
\left(
1-\frac{\sqrt{(R^{2}-r_{+}^{2})(R^{2}-r_{-}^{2})}}{R^{2}}
\right),
\qquad
J=\frac{r_{+}r_{-}}{4G\ell},
\label{eq:sec15_EJ}
\end{equation}
\begin{equation}
T_{R}
=
\frac{R(r_{+}^{2}-r_{-}^{2})}{2\pi\ell r_{+}\sqrt{(R^{2}-r_{+}^{2})(R^{2}-r_{-}^{2})}},
\qquad
\Omega_{R}
=
\frac{r_{-}\sqrt{R^{2}-r_{+}^{2}}}{r_{+}R\sqrt{R^{2}-r_{-}^{2}}},
\label{eq:sec15_TROR}
\end{equation}
satisfy the exact first law
\begin{equation}
dE=T_{R}\,dS+\Omega_{R}\,dJ-P\,dL,
\qquad
L=2\pi R.
\label{eq:sec15_firstlaw}
\end{equation}
The static branch is canonically stable throughout \(0<r_{+}<R\), while the rotating grand potential has a positive-definite Hessian throughout the non-extremal physical domain. The cavity therefore defines a finite thermodynamic system whose local response is completely controlled by wall variables.

The global phase structure of the static ensemble is determined by the competition between Euclidean BTZ and thermal AdS\(_3\) at fixed wall torus. The transition occurs at
\begin{equation}
T_{c}(R)=\frac{1}{2\pi R},
\label{eq:sec15_Tc}
\end{equation}
so the critical temperature is fixed entirely by the proper wall radius. In wall units this is the statement that the phase boundary occurs when the proper thermal cycle and the spatial cycle of the wall torus have equal length. The Hawking--Page transition thus becomes a finite-size phase transition of the cutoff theory defined on a circle of circumference \(2\pi R\). This is one of the sharpest global consequences of the finite-cutoff viewpoint because the phase boundary is determined directly by the geometry of the wall rather than by asymptotic data \cite{HawkingPage1983,Witten1998Thermal,MaloneyWitten2010,HuangTao2022}.

A second central result concerns radial flow. Keeping the bulk state fixed while moving the wall generates exact first-order flow equations for the wall observables,
\begin{equation}
R\left(\frac{\partial \epsilon}{\partial R}\right)_{r_{+},r_{-}}
=
-(\epsilon+p),
\qquad
R\left(\frac{\partial j}{\partial R}\right)_{r_{+},r_{-}}
=
-2j,
\label{eq:sec15_radialflow}
\end{equation}
together with the corresponding flows of \(T_{R}\) and \(\Omega_{R}\). The wall radius therefore acts simultaneously as the physical size of the quasilocal system and as the running scale of the finite-cutoff theory. In the ultraviolet limit \(R\gg \ell,r_{+},r_{-}\), the trace vanishes and the wall stress tensor approaches the ordinary conformal equation of state. Moving the wall inward strengthens the finite-cutoff deformation while preserving the exact closure of the wall thermodynamics.

The microscopic interpretation follows the same logic. The ultraviolet cylinder theory supplies the state-counting through Cardy growth, while the cavity supplies the exact nonlinear map from ultraviolet charges to wall observables. In the static sector this map is
\begin{equation}
M=\frac{R}{\ell}E-2GE^{2},
\label{eq:sec15_ME}
\end{equation}
so the entropy becomes the deformed Cardy formula
\begin{equation}
S(E,R)
=
2\pi
\sqrt{
\frac{c}{3}
\left(
RE-2G\ell E^{2}
\right)
},
\qquad
c=\frac{3\ell}{2G}.
\label{eq:sec15_deformedCardy}
\end{equation}
In the rotating sector the entropy retains the exact chiral Cardy form in terms of the undeformed ultraviolet energies \(E_{0}^{\pm}\), while the map between \((E,\Pi)\) and \((E_{0},\Pi)\) is deformed by the finite cutoff. The one-loop correction refines this picture through the solid-torus determinant evaluated on the wall modulus and the Gaussian susceptibility factor from the inverse Laplace--Fourier transform. The finite-cutoff density of states is therefore the microscopic counterpart of the quasilocal thermodynamics rather than an independent construction \cite{Cardy1986,Strominger1998,MaloneyWitten2010}.

The picture that emerges is conceptually rigid. The cavity is neither a regulator to be removed nor a merely auxiliary boundary condition. It is the finite holographic screen on which the quasilocal stress tensor, the finite-radius thermodynamics, the radial Hamilton--Jacobi flow, and the deformed microscopic spectrum are all realized by the same exact wall system. The finite-cutoff formulation is therefore strongest where it is most constrained: in AdS\(_3\) Einstein gravity, where the absence of local bulk degrees of freedom allows the entire construction to close on the renormalized wall observables.

Several extensions are natural. A first direction is the full rotating phase diagram at fixed \((R,T_{R},\Omega_{R})\), where one expects a finite-cutoff version of modular competition on a twisted torus. A second direction is the inclusion of gauge fields or scalar hair, which would enlarge the wall dictionary by introducing additional quasilocal chemical potentials and modified flow equations. A third direction is higher-derivative gravity, where the Brown--York tensor is replaced by the appropriate generalized boundary stress tensor and the entropy by the Wald or Dong entropy, giving a direct route to finite-cutoff deformations beyond pure Einstein gravity. A fourth direction is the exact modular structure of the full finite-cutoff partition function, which should connect the semiclassical wall-torus analysis to deformed torus amplitudes in solvable two-dimensional theories. A fifth direction is nonequilibrium dynamics, where radial flow may provide a natural geometric description of finite-cutoff transport, relaxation, and response.

The overarching lesson of the paper is that BTZ black holes in a cavity provide an exactly tractable setting in which quasilocal thermodynamics, Euclidean saddle analysis, holographic RG flow, finite-cutoff AdS\(_3\)/CFT\(_2\), and microscopic state counting meet in one coherent framework. The cavity is therefore best understood as a finite holographic screen whose thermodynamics encodes the scale dependence of the dual theory in a concrete and computable way.

\appendix

\section{Details of the static Euclidean action}
\label{app:static_action}

This appendix collects the standard Euclidean-action algebra used in Section~5. For the BTZ saddle
\begin{equation}
ds^2=f(r)d\tau^2+\frac{dr^2}{f(r)}+r^2d\phi^2,
\qquad
f(r)=\frac{r^2-r_+^2}{\ell^2},
\qquad
r\in[r_+,R],
\end{equation}
one has \(\mathcal R=-6/\ell^2\), \(\sqrt g=r\), and \(\beta_H=2\pi\ell^2/r_+\). The bulk contribution is
\begin{equation}
I^{\rm BH}_{\rm bulk}
=
\frac{\beta_H}{4G\ell^2}(R^2-r_+^2).
\label{eq:app_IbulkBH}
\end{equation}
At the wall,
\begin{equation}
K(R)=\frac{f'(R)}{2\sqrt{f(R)}}+\frac{\sqrt{f(R)}}{R}
=
\frac{R}{\ell\sqrt{R^2-r_+^2}}+
\frac{\sqrt{R^2-r_+^2}}{\ell R},
\qquad
\sqrt h=\frac{R\sqrt{R^2-r_+^2}}{\ell}.
\label{eq:app_KBH}
\end{equation}
Thus
\begin{equation}
I^{\rm BH}_{\rm GHY}
=-\frac{\beta_H}{4G\ell^2}(2R^2-r_+^2),
\qquad
I^{\rm BH}_{\rm ct}
=
\frac{\beta_HR}{4G\ell^2}\sqrt{R^2-r_+^2}.
\label{eq:app_IGHYIctBH}
\end{equation}
Adding the three terms gives
\begin{equation}
I_{\rm BH}(R,r_+)
=
\frac{\pi R}{2Gr_+}\left(\sqrt{R^2-r_+^2}-R\right).
\label{eq:app_IBH}
\end{equation}

For thermal AdS\(_3\),
\begin{equation}
ds^2=f_0(r)d\tau^2+\frac{dr^2}{f_0(r)}+r^2d\phi^2,
\qquad
f_0(r)=1+\frac{r^2}{\ell^2},
\qquad
r\in[0,R].
\end{equation}
The wall period is
\begin{equation}
\beta_0=\frac{\beta_R}{\sqrt{1+R^2/\ell^2}}.
\label{eq:app_beta0}
\end{equation}
Evaluation of the same bulk, Gibbons--Hawking, and counterterm pieces yields
\begin{equation}
I_{\rm AdS}(R,\beta_R)
=
\frac{\beta_R}{4G\ell}\left(R-\sqrt{R^2+\ell^2}\right),
\qquad
F_{\rm AdS}(R)=\frac{1}{4G\ell}\left(R-\sqrt{R^2+\ell^2}\right).
\label{eq:app_IAdS}
\end{equation}

\bibliographystyle{ytphys.bst}
\bibliography{references}

\end{document}